\documentclass[english, 12pt]{article}
\usepackage[T1]{fontenc}
\usepackage[utf8]{inputenc}
\usepackage[top=1.25in, bottom=1.25in, left=1.25in, right=1.25in]{geometry}
\usepackage[pdftex]{graphicx}
\usepackage{verbatim}
\usepackage[width=0.8\textwidth]{caption}
\usepackage{amsmath,amssymb,amsopn,dsfont,mathrsfs, bm}
\usepackage[english]{babel}
\usepackage{lmodern}
\usepackage{flafter, float,placeins,longtable,array}
\usepackage{natbib}
\usepackage{euscript}
\usepackage{color}
\usepackage{lscape,subfig,threeparttable}
\usepackage{adjustbox}

\usepackage{pdflscape}
\usepackage{afterpage}

\usepackage{multirow}
\usepackage{enumitem}
\usepackage{booktabs}
\usepackage{stmaryrd}
\usepackage[hidelinks]{hyperref}
\definecolor{navyblue}{rgb}{0.0, 0.0, 0.5}
\hypersetup{colorlinks=true, citecolor=navyblue, linkcolor=black, urlcolor=navyblue}
\makeatletter
\newtheorem{thm}{Theorem}[section]

\newtheorem{lem}[thm]{Lemma}
\newtheorem{prop}[thm]{Proposition}
\newtheorem{hyp}{Assumption}

\numberwithin{equation}{section}

\newcommand{\R}{\mathbb R}
\renewcommand{\P}{\mathbb P}
\newcommand{\E}{\mathbb{E}}
\newcommand{\V}{\mathbb{V}}
\newcommand{\eps}{\varepsilon}
\newcommand{\sgn}{\text{sgn}}
\newcommand{\lin}{\text{lin}}
\newcommand{\Supp}{\text{Supp}}

\newcommand{\indep}{\perp \!\!\! \perp}

\newcommand{\Deriv}[2]{\frac{\partial #1}{\partial #2}}

\renewcommand{\triangleq}{:=}
\newcommand{\ind}[1]{\mathds{1}\left\{#1\right\}}

\renewcommand{\section}{\@startsection{section}{2}{0mm}{-1.5\baselineskip}{1\baselineskip}{\normalfont\Large\bfseries}}
\renewcommand{\subsection}{\@startsection{subsection}{2}{0mm}{-1.2\baselineskip}{1\baselineskip}{\normalfont\normalsize\bfseries}}
\renewcommand{\subsubsection}{\@startsection{subsubsection}{3}{0mm}{-0.8\baselineskip}{0.4\baselineskip}{\normalfont\normalsize\itshape}}

\date{}

\begin{document}
	
	\title{Fixed Effects Binary Choice Models with Three or More Periods\thanks{We would like to thank Pascal Lavergne, the co-editor Limor Golan and three anonymous referees for their helpful comments.}}
	\author{Laurent Davezies\thanks{CREST-ENSAE, laurent.davezies@ensae.fr} \and Xavier D'Haultf\oe uille\thanks{CREST-ENSAE, xavier.dhaultfoeuille@ensae.fr. Xavier D'Haultf\oe uille thanks the hospitality of PSE where part of this research was conducted. He also gratefully acknowledges financial support from the research grants Otelo (ANR-17-CE26-0015-041).} \and Martin Mugnier\thanks{CREST-ENSAE, martin.mugnier@ensae.fr.}}
	
	\maketitle
	
	\begin{abstract}
We consider fixed effects binary choice models with a fixed number of periods $T$ and regressors without a large support. If the time-varying unobserved terms are i.i.d. with known distribution $F$, \cite{chamberlain2010} shows that the common slope parameter is point identified if and only if $F$ is logistic. However, he only considers in his proof $T=2$. We show that the result does not generalize to $T\geq 3$: the common slope parameter can be identified when $F$ belongs to a family including the logit distribution. Identification is based on a conditional moment restriction. Under restrictions on the covariates, these moment conditions lead to point identification of relative effects. If $T=3$ and mild conditions hold, GMM estimators based on these conditional moment restrictions reach the semiparametric efficiency bound. Finally, we illustrate our method by revisiting \cite{BrenderDrazen2008}.

\end{abstract}
	
\newpage
		
\section{Introduction}
	
In this paper, we revisit the classical binary choice model with fixed effects. Specifically, let $T$ denote the number of periods and let us suppose to observe, for individual $i$, $(Y_{it}, X_{it})_{t=1, \dotsc, T}$ with
	\begin{equation}\label{eq:base_line_panel_binary_model}
	Y_{it} = \mathds{1}\{X_{it}'\beta_0 \, + \, \gamma_i - \varepsilon_{it} \geq 0\}
	\end{equation}
where $\beta_0\in \R^K$ is unknown and  $\varepsilon_{it} \in \R$ is an idiosyncratic shock. The nonlinear nature of the model and the absence of restriction on the distribution of $\gamma_i$ conditional on $X_i:=(X_{i1}',\dotsc,X_{iT}')'$ renders the identification of $\beta_0$ difficult. \cite{rasch1960} shows that if the $(\eps_{it})_{t=1, \dotsc, T}$ are i.i.d. with a logistic distribution, a conditional maximum likelihood can be used to identify and estimate $\beta_0$.  \cite{chamberlain2010} establishes a striking converse of Rasch's result: if the $(\eps_{it})_{t=1, \dotsc, T}$ are i.i.d. with distribution $F$ and the support of $X_i$ is bounded, $\beta_0$ is point identified only if $F$ is logistic. Other papers have circumvented such an impossibility result by either considering large support regressors \citep[see in particular][]{manski1987, HonoreLewbel2002} or allowing for dependence between the shocks \citep[see][]{Magnac2004}.
	
\medskip
	It turns out, however, that \cite{chamberlain2010} only proves his result for $T=2$.  And in fact, we show that his result does not generalize to $T\geq 3$. Specifically, we consider distributions $F$ satisfying
	\begin{equation}
	\label{eq:cond_OR}
	\frac{F(u)}{1-F(u)}=\sum_{k=1}^{\tau} w_k \exp(\lambda_{k} u) \; \text{ or } \frac{1-F(u)}{F(u)}=\sum_{k=1}^{\tau} w_k \exp(-\lambda_{k} u),
	\end{equation}
	with $T\geq \tau+1$, $(w_1,...,w_\tau) \in (0,\infty)^\tau$ and $1=\lambda_{1} < \dotsc < \lambda_{\tau}$. We study the identification of $\beta_0$, assuming that $\lambda:=(\lambda_{1},\dotsc,\lambda_{\tau})$ is known. The weights $w_1,\dotsc,w_{\tau}$ remain unknown, thus allowing for much more flexibility on the distribution of $\eps_{it}$ than in the logit case. In particular, it may either be left- or right-skewed, platykurtic or leptokurtic.  Our main insight is that for any $F$ satisfying \eqref{eq:cond_OR}, a conditional moment restriction holds. We also obtain some results on the corresponding identified set $B$. For instance, if, roughly speaking, $X_i$ is continuous, we show that $B$ includes at most $T!-1$ points (2 if $T=3$) and relative marginal effects are point identified. Note that \cite{johnson_2004} considers the same family with $\tau=2$ and $T=3$. However, he does not study the general case and does not show any formal identification result based on the corresponding moment conditions.
	
\medskip
Obviously, the conditional moment condition can be used to construct GMM estimators. This means, in particular, that $\sqrt{n}$-consistent estimation is possible beyond the logit case when $T>2$, overturning again the impossibility results of \cite{chamberlain2010} and \cite{Magnac2004}. Further, we show that if $T=3$ and mild additional restrictions hold, the optimal GMM estimator based on our conditional moment conditions reaches the semiparametric efficiency bound of the model. Hence, at least when $T=3$, these moment conditions contain all the information of the model.

\medskip
Finally, we showcase the empirical relevance of our approach by studying whether budget deficits and economic growth affect reelection, revisiting \cite{BrenderDrazen2008}. The authors investigate this issue using simple and fixed effects logit models.
However, the assumption of logistic errors is not warranted, so we consider whether the results are robust to this assumption on the unobserved terms. Our results suggest that the relative effects of budget deficits and economic growth or other variables are fairly robust to the logistic assumption.

\medskip
Our paper is related to the seminal work of \cite{Bonhomme_2012}, who develops a unified approach for models where the conditional distribution of $(Y_1,...,Y_T)$ given $(X_i,\gamma_i)$ is parametrized by $\beta_0$, but no restriction on the distribution of $\gamma_i|X_i$ is imposed. In such set-ups, he shows that the identification and estimation of $\beta_0$ depends on the existence of functions $m\neq 0$ satisfying
$$\E(m(Y,X,\beta_0)|X,\gamma)=0.$$
This approach has been fruitfully applied to the dynamic logit model by \cite{Kitazawa2022} and \cite{honore2020}. Our paper may be seen as yet another application of this approach, focusing on static models but dropping the logistic assumption.

\medskip
The remainder of the paper is organized as follows. Section~\ref{sec:identification} describes the moment condition we use for identification of $\beta_0$ and establishes some properties of the identified set based on these moments. Section~\ref{sec:estimation} discusses GMM estimation of $\beta_0$, links it with the semiparametric efficiency bound of the model and discusses the case of unbalanced panel data. Section~\ref{sec:appli} is devoted to the application. Section~\ref{sec:conclusion} concludes. All the proofs are collected in the appendix.

\section{Identification}\label{sec:identification}
	
\subsection{The model and moment conditions} 
\label{sub:the_model_and_moment_conditions}
	
We drop the subscript $i$ in the absence of ambiguity and let $Y=(Y_{1}',\dotsc, Y_{T}')'$, $X=(X_{1}',\dotsc, X_{T}')'$, $X_t=(X_{1,t},\dotsc ,X_{K,t})'$ , $X_{k,\cdot}=(X_{k,1}, \ldots, X_{k,T})'$, $X_{-k}= (X_{k',t})_{k'\neq k, t=1, \dotsc, T}$, $X_{k,-t}=(X_{k,s})_{s\neq t}$, and $X_{-k,t}=(X_{k',t})_{k'\neq k}$. $\Supp(X)\subset \R^{KT}$ denotes the support of the random variable $X$. For any set $A\subset \R^p$ (for any $p\geq 1$), we let $A^*:=A\backslash\{0\}$ and denote by $|A|$ the cardinal of $A$. Hereafter, we maintain the following conditions.

	\begin{hyp}[Binary choice panel model]\label{as:basic}
		Equation~(\ref{eq:base_line_panel_binary_model}) holds and:
		\begin{enumerate}
			\item $(X, \gamma)$ and $(\varepsilon_{t})_{1\leq t\leq T}$ are independent and the $(\varepsilon_{t})_{1\leq t\leq T}$ are i.i.d. with a known cumulative distribution function (cdf) $F$.
			\item For all $(k,t)$, $\E[X_{k,t}^2]<\infty$.
			\item $\beta_0 \in \R^{K*}$.
		\end{enumerate}
	\end{hyp}
	
The first condition is also considered in \cite{chamberlain2010}. The second condition is a standard moment restriction on the covariates. Finally, we exclude in the third condition the case $\beta_0=0$ here. This case can be treated separately, as the following proposition shows.
	
	\begin{prop}\label{prop:beta0_0}
		Suppose that Assumption~\ref{as:basic} holds, $F$ is strictly increasing on $\R$ and there exist $(t,t')\in\{1,\dotsc, T\}^2$ such that $\E[(X_t-X_{t'})(X_t-X_{t'})']$ is nonsingular. Then $\beta_0=0$ if and only if
		\begin{equation}
		\label{eq:test_beta0_0}
		\P(Y_t=1,Y_{t'}=0|Y_t+Y_{t'}=1, X_t, X_{t'})=\frac{1}{2} \quad \text{a.s.}
		\end{equation}
	\end{prop}
	
	Condition~\eqref{eq:test_beta0_0} can be tested by a specification test on the nonparametric regression of $D=Y_t(1-Y_{t'})$ on $(X_t,X_{t'})$, conditional on the event $Y_t+Y_{t'}=1$. See, e.g., \cite{bierens1990consistent} or \cite{hong1995consistent}.
	
\medskip
Turning to identification on $\R^{K*}$, we first recall the impossibility result of \cite{chamberlain2010}. We say below that $F$ is logistic if $G(u):=F(u)/(1-F(u))=w\exp(\lambda u)$ for some $(w,\lambda)\in\R^{+*2}$.
	
\begin{thm}\label{thm:chamberlain}
	Suppose that $T=2$, $X_t$ includes $\mathds{1}\{t=2\}$, Assumption~\ref{as:basic}.1 holds, $F$ is strictly increasing on $\R$ with bounded, continuous derivative and $\Supp(X)$ is compact. If $F$ is not logistic, there exists $\beta_0 \in \R^{K*}$, a distribution of $\gamma|X$ and an open ball $B\subset \R^K$ such that $\beta_0$ is not identified compared to $\beta\in B$.
\end{thm}

This result implies in particular that when $T=2$ and $F$ is not logistic, relative effects $\beta_{0j}/\beta_{0k}$, for $k$ such that $\beta_{0k}\neq 0$, may not be identified. Such relative effects are important as they are equal to relative marginal effects if both $X_{j,t}$ and $X_{k,t}$ are continuous. If only $X_{k,t}$ is continuous (say), $-\beta_{0j}/\beta_{0k}$ still corresponds to a compensating variation.\footnote{To see the first point, note that under Assumptions \ref{as:basic}-\ref{as:gen_logit},
$$\mu_{k,t}(x):=\Deriv{\P(Y_t=1|X_{k,t} = x_{k,t}, X_{k,-t} = x_{k,-t} , X_{-k}=x_{-k})}{x_{k,t}} = \beta_{0k}\E[F'(x_t'\beta_0 + \gamma)|X=x]$$
and thus $\mu_{j,t}(x)/\mu_{k,t}(x) = \beta_{0j}/\beta_{0k}$. Also, $-\beta_{0j}/\beta_{0k}$ corresponds to the change in $X_{k,t}$ necessary to keep $\P(Y_t=1|X_t,\alpha)$ constant when $X_{j,t}$ increases by one unit.}

\medskip
The key step in Chamberlain's proof is that if $\beta_0$ is identified for all data generating process satisfying the restrictions of the theorem, the conditional probabilities (conditional on $X$ and $\gamma$) of the four possible trajectories for $(Y_1,Y_2)$ are necessarily affinely dependent. Moreover, by letting $|\gamma|$ tend to infinity, the stable trajectories $(0,0)$ and $(1,1)$ disappear from this relationship. This leads to the following functional equation for $G$:
\begin{align}\psi_1(\alpha) G(u) + \psi_2(\alpha) G(u+\alpha)=0,\label{eq:fonc_T2} \end{align}
for all $u\in \R$, $\alpha$ in an open subset of $\R$ and some functions $\psi_1(\cdot),\psi_2(\cdot)$ such that for all $\alpha$, $(\psi_1(\alpha),\psi_2(\alpha))\neq (0,0)$. The result follows by noting that the solutions necessarily have the form $u\mapsto w\exp(\lambda u)$.

\medskip
Equation \eqref{eq:fonc_T2} relies on the time dummy variable $\ind{t=2}$. However, the proof of Theorem 2 of \cite{chamberlain2010} shows that even without such a dummy variable, \eqref{eq:fonc_T2} is necessary for the semiparametric efficiency bound not to be zero, or, equivalently, for the existence of regular, root-n consistent estimators of $\beta_0$. In this case, $\alpha$ corresponds to $(x_2-x_1)'\beta_0$, for $(x_1,x_2)$ in a set of positive measure.

\medskip
In any case, the same reasoning with $T=3$ leads to the following equation for $G$:
\begin{align}
\psi_1(\bm{\alpha}) G(u) + \psi_2(\bm{\alpha}) G(u+\alpha_1)+ \psi_3(\bm{\alpha}) G(u+\alpha_2) + \psi_4(\bm{\alpha}) G(u) G(u+\alpha_1) & \notag \\
+ \psi_5(\bm{\alpha})
G(u) G(u+\alpha_2)+ \psi_6(\bm{\alpha}) G(u+\alpha_1) G(u+\alpha_2)&=0,	
	\label{eq:fonc_T3}
\end{align}
for all $u\in \R$, $\bm{\alpha}:=(\alpha_1,\alpha_2)$ in an open subset of $\R^2$ and some functions $\psi_k(\cdot)$, $k=1,...,6$, such that for for all $\bm{\alpha}$, $(\psi_1(\bm{\alpha}),...,\psi_6(\bm{\alpha}))\neq (0,...,0)$. We now have $6=2^3 - 2$ terms instead of just $2=2^2-2$, and thus we can expect to have other solutions than just $u\mapsto w\exp(\lambda u)$. And indeed, one can check that if $G$ has the form $u\mapsto w_1\exp(\lambda_1 u)+w_2\exp(\lambda_2 u)$, we can construct $(\psi_1(\bm{\alpha}),\psi_2(\bm{\alpha}),\psi_3(\bm{\alpha}))\neq (0,0,0)$ such that \eqref{eq:fonc_T3} holds, with $\psi_4(\bm{\alpha})=\psi_5(\bm{\alpha})=\psi_6(\bm{\alpha})=0$. Similarly, if $1/G$ has the form $u\mapsto w_1\exp(\lambda_1 u)+w_2\exp(\lambda_2 u)$, we can construct $(\psi_4(\bm{\alpha}),\psi_5(\bm{\alpha}),\psi_6(\bm{\alpha}))\neq (0,0,0)$ such that \eqref{eq:fonc_T3} holds, with $\psi_1(\bm{\alpha})=\psi_2(\bm{\alpha})=\psi_3(\bm{\alpha})=0$. Note that there may still be other solutions to \eqref{eq:fonc_T3} that are increasing and have a limit of $\infty$ (resp. $0$) at $\infty$ (resp. at $-\infty$). The question of identifying all such solutions is left for future research.

\medskip
Generalizing this reasoning to any $T>2$, we see that combinations of at most $T-1$ exponential functions satisfy the functional restrictions tantamount to \eqref{eq:fonc_T3} and which render identification of $\beta_0$ possible. This suggests that identification may be achieved for the corresponding family of distribution, which we now formally introduce. Hereafter, $\Lambda_{\tau}$ denotes a subset of $\{(\lambda_1,\dotsc, \lambda_{\tau})\in\R^{\tau}: 1=\lambda_1<\dotsc< \lambda_{\tau}\}$.
	
	\begin{hyp}[``Generalized'' logistic distributions]\hspace{-0.2cm}\footnote{Though we use the same name, our family of distributions should not be confused with those introduced by \cite{balakrishnan1988order} and \cite{stukel1988generalized}.}\label{as:gen_logit}
		There exist known $\tau \in \{1, \ldots, T-1\}$ and $\lambda \triangleq (\lambda_{1}, \dotsc, \lambda_{\tau})' \in \Lambda_{\tau}$ and unknown $w \triangleq (w_1, \dotsc, w_{\tau})' \in (0,\infty)^\tau$ such that:
		$$\begin{array}{rccl}
		\text{Either } & F(u)/(1-F(u)) =& \sum_{j=1}^{\tau} w_j \exp(\lambda_{j}u) & \text{(First type)}, \\[3mm]
		\text{or } & (1-F(u))/F(u) =& \sum_{j=1}^{\tau} w_j \exp(-\lambda_{j}u) & \text{(Second type)}.	
		\end{array}$$
	\end{hyp}
	
\medskip
Fixing $\min\{\lambda_{1},\dotsc, \lambda_{\tau}\}$ to 1 is without loss of generality, as we can always multiply $\beta_0$, $\gamma_i$ and $\eps_{it}$ by this factor. If $F$ is of the second type, then one can show that the cdf of $-\eps_{it}$ is of the first type. Thus, up to changing $(Y_{t},X_{t})$ into $(1-Y_{t},-X_{t})$, we can assume without loss of generality, as we do afterwards, that $F$ is of the first type. We shall see that $\tau+1$ periods are sufficient to achieve identification. Hence, we assume, again without loss of generality, that $T=\tau+1$: if $T>\tau+1$, we can always focus on  $\tau+1$ periods.
	
\medskip
Before describing our identification strategy of $\beta_0$ when $F$ is a generalized logistic distribution, two remarks are in order. First, we obtain our results below irrespective of the vector $w$.\footnote{\label{foot:w_zero} We do impose however that all the components of $w$ are non-zero, for normalization purposes. Otherwise, the model with $w=(w_1,0)$ and $\beta_0$, for instance, would be equivalent to the model with $w=(0,w_1)$ and $\beta_0/\lambda_2$. A similar issue arises  with, e.g., $w=(w_1,w_2,0)$ if $\lambda_3/\lambda_2=\lambda_2/\lambda_1$.}  Hence, in contradistinction with the fixed effect logistic model, we do not fix the distribution of $\eps$, but simply impose that it belongs to a family of distributions indexed by two parameters. Members of this family differ in particular by their skewness and kurtosis. In linear regressions, the residuals are often found to have a skewed distribution with either positive or negative excess kurtosis. Then, there is no reason why the latent variables corresponding to $Y_{it}$ would not exhibit a similar pattern. On the other hand, we do fix $\lambda$. Identification of $\lambda$ could also be of interest but is not addressed in this paper.

\medskip
Now, the idea behind the identification of $\beta_0$ is to construct a function $m\neq 0$ such that
$\E(m(Y,X,\beta_0)|X,\gamma)=0$ almost surely. Thus, as mentioned in the introduction, we apply \cite{Bonhomme_2012}'s general idea of functional differencing. The function $m$ is related to the functions $\psi_k$ in \eqref{eq:fonc_T3} when $T=3$, and the generalization of \eqref{eq:fonc_T3} when $T>3$. For any $x=(x'_1,...,x'_T)'\in\R^{KT}$, let $x_s^{-t}{}=x_s$ if $s<t$,  $x_s^{-t}{}=x_{s+1}$ else. We let
$$M_t(x;\beta)= (-1)^{t+1} \det\begin{pmatrix}
	\exp(\lambda_{1}x_1^{-t}{}'\beta) & \ldots & \exp(\lambda_{1}x_{T-1}^{-t}{}'\beta) \\
	\vdots &  & \vdots \\
	\exp(\lambda_{T-1}x_1^{-t}{}'\beta) & \ldots & \exp(\lambda_{T-1}x_{T-1}^{-t}{}'\beta)
\end{pmatrix}.$$
Then define, for any $(y,x,\beta)\in\{0,1\}^T\times \Supp(X)\times \R^{K*}$,
$$m(y, x; \beta):=\sum_{t=1}^T \mathds{1}\{y_t=1, y_{t'}=0 \;\forall t'\neq t\} M_t(x;\beta).$$
Our first result shows that $m$, indeed, satisfies a conditional moment restriction:
	
\begin{thm}\label{thm:moments}
	If Assumptions~\ref{as:basic}-\ref{as:gen_logit} hold, we have,  almost surely,
	\begin{equation}
	\E[m(Y,X;\beta_0) | X,\gamma]=\E[m(Y,X;\beta_0) | X]=0.	
		\label{eq:condit_mom}
	\end{equation}
\end{thm}
	
Theorem~\ref{thm:moments} shows there exists a known moment condition which potentially identifies $\beta_0$ in a more general model than the logistic one. Also, as the number of periods $T$ increases, the class of distributions $F$ for which $\beta_0$ can be point identified increases. This is consistent with the idea that if $T=\infty$, $\beta_0$ is point identified for any $F$, by using variations in $X_t$ of a single individual. Note however that the class of generalized logistic distribution is not dense for the set of all cdf's: any cdf $F$ belonging to the closure of this class should be such that either $F/(1-F)$  or $(1-F)/F$ is convex. Theorem~\ref{thm:moments}  also complements the results of \cite{chernozhukov2013average} showing that bounds on $\beta_0$ for general $F$ shrink quickly as $T$ increases.

\medskip
Theorem~\ref{thm:moments} holds with $T=\tau+1=2$. In such a case, the conditional moment condition can be written
$$\E\left[\mathds{1}\{Y_1>Y_2\} \exp(X'_2\beta_0)- \mathds{1}\{Y_2>Y_1\}  \exp(X'_1\beta_0)|X\right]=0.$$
This conditional moment generates the first-order conditions of the maximization of the theoretical conditional likelihood, since these the first-order conditions are equivalent to
$$\E\left[\frac{(X_1-X_2)}{\exp(X_1'\beta_0) + \exp(X_2'\beta_0)}\left(\mathds{1}\{Y_1>Y_2\} \exp(X'_2\beta_0)- \mathds{1}\{Y_2>Y_1\}  \exp(X'_1\beta_0)\right)\right]=0.$$


\subsection{Necessary and sufficient conditions for identification} 
\label{sub:necessary_and_sufficient_conditions_for_identification}

The discussion above implies that with $T=\tau+1=2$, $\beta_0$ is identified by \eqref{eq:condit_mom} as soon as $\E\left[(X_1-X_2)(X_1-X_2)'\right]$ is nonsingular. We now turn to the more difficult case where $T-1=\tau>1$. Let $B$ denote the identified set of $\beta_0$ obtained with our conditional moment conditions, namely
$$B := \left\{b \in \R^{K*} : \E[m(Y,X;b)|X] = 0 \; \text{a.s.}\right\}.$$
We also denote by $B_k:=\{b_k:\exists b=(b_1,...,b_k,...,b_K)\in B\}$ ($k=1,...,K$) the identified set of $\beta_{0k}$. Our first result shows that $B$ is included in a set depending on the distribution of $X$ only. To define this set, let us introduce
\begin{align*}
	D_j(x;b) & := \det\begin{pmatrix} \exp(\lambda_{j}x_1'\beta_0) & \ldots & \exp(\lambda_{j}x_T'\beta_0) \\
		\exp(\lambda_{1}x_1'b) & \ldots & \exp(\lambda_{1}x_T'b) \\
		\vdots & & \\
		\exp(\lambda_{T-1} x_1'b) & \ldots & \exp(\lambda_{T-1} x_T'b)
	\end{pmatrix}
\end{align*}
and, for all $b\in\R^{K*}$, let
\begin{align*}
	\mathcal{D}(b)  =&  \left\{x \in \Supp(X): \max_{j=1,...,T-1} D_j(x;b)> \min_{j=1,...,T-1} D_j(x;b) \geq 0 \right. \\
	& \hspace{2.2cm} \left. \text{ or }  \min_{j=1,...,T-1} D_j(x;b)< \max_{j=1,...,T-1} D_j(x;b) \leq 0\right\}.
\end{align*}
Because $D_j(x;\beta_0)=0$ for all $x \in \Supp(X)$, we have $\P(X \in \mathcal{D}(\beta_0))=0$. The following lemma shows that $B$ is actually included in the set of $b$'s satisfying this property.

\medskip
\begin{lem}\label{lem:high_level_id}
	Suppose that Assumptions \ref{as:basic}-\ref{as:gen_logit} hold. Then,
	$$B\subset \widetilde{B}:=\left\{b\in \R^{K*}:\; \P(X \in \mathcal{D}(b))=0\right\}.$$
\end{lem}

This result follows because the moment condition can be written as a weighted sum of the $D_j(x;b)$'s, with positive weights. It shows that $\beta_0$ is identified if for all nonzero $b \neq \beta_0$, we can find some $x\in\Supp(X)$ such that all nonzero $D_j(x;b)$ have the same sign, and the set of such nonzero determinants is not empty.

\medskip
The set $\widetilde{B}$ is convenient in that it does not depend on the unknown distribution of $\gamma|X$; but it is hard to characterize in general. Nevertheless, we are able to obtain results under either of the conditions below.

\begin{hyp}\label{as:zero_in_supp}
For all $k\in\{1,...,K\}$,  $\P\left(|\left\{X_{k,1},...,X_{k,T}\right\}|=T,X_{-k}=0\right)>0$.\footnote{When $K=1$, the condition $X_{-k}=0$ should simply be omitted.}	
\end{hyp}

\begin{hyp}\label{as:cont_regressors}
There exists $(s,t,x)\in \{1,…,T\}^2 \times \R^K$, $s< t$ and a neighborhood $V$ of $x$ such that $\Supp(X)\cap [\R^{(s-1)K} \times V \times \R^{(t-s-1)K} \times V \times \R^{(T-t)K}]$ has a non-empty interior.
\end{hyp}

The first assumption corresponds to a case where all components of $X$ are discrete. It imposes that for all $k$ and $t$, the support of $X_{k,t}$ includes 0 and at least $T-1$ additional elements. Because we can always replace $X_{k,\cdot}$ by $X_{k,\cdot}-c_k$ for any $c_k\in\R^T$, the condition $0\in\Supp(X_{k,t})$ for all $k, t$ holds as long as $\cap_{t=1}^T \Supp(X_{k,t})$ is not empty (for all $k$). The second condition imposes that all components of $X_t$ are continuous. It also imposes that for at least two periods $s$ and $t$, $\Supp(X_s)\cap \Supp(X_t)$ is not empty. This last condition holds for instance if $(X_t)_{t\geq 1}$ is strictly stationary.

\begin{thm}\label{thm:main_continuous}
	Suppose that Assumptions \ref{as:basic}-\ref{as:gen_logit} hold. Then:
\begin{enumerate}
	\item If Assumption \ref{as:zero_in_supp} also holds, $|B|<\infty$ and $B_k \subset \{c\beta_{0k}: c \in \{0\}\cup (1/\lambda_{T-1},\lambda_{T-1})\}$.
	\item If Assumption \ref{as:cont_regressors} also holds,
	\begin{equation}
B \subset \widetilde{B} \subset R:=\{c\beta_0 : c \in  (1/\lambda_{T-1}, \lambda_{T-1})\}.		
		\label{eq:caract_B}
	\end{equation}
	Moreover, $|B|\leq T!-1$ and $|B |\leq 2$ when $T=3$. All relative effects $\beta_{0j}/\beta_{0k}$, for $k$ such that $\beta_{0k}\neq 0$, are point identified.\footnote{The set of indices $k$ such that $\beta_{0k}\neq 0$ is identified since by \eqref{eq:caract_B}, $B_k=\{0\}$ when $\beta_{0k}=0$, and $0\not\in B_k$ otherwise.}
\end{enumerate}	
\end{thm}

Whether Assumption \ref{as:zero_in_supp} or \ref{as:cont_regressors} holds, Theorem \ref{thm:main_continuous} shows that under-identification is at most finite, namely $|B|<\infty$. This implies that $\beta_0$ is locally identified in the sense that there exists a neighborhood of $\beta_0$ in which the unique solution to the equation $\E[m(Y,X;b) | X]=0$ is $b=\beta_0$. Further, the first result of Theorem \ref{thm:main_continuous} shows that with discrete regressors satisfying Assumption \ref{as:zero_in_supp}, the ``length'' of the identified set on $\beta_{0k}$, defined as
$$\max_{(b_{1k},b_{2k})\in B_k^2} |b_{1k}-b_{2k}|,$$
cannot exceed $\beta_{0k}(\lambda_{T-1}-1/\lambda_{T-1})$ if $0\not\in B_k$. Note that under Assumption \ref{as:zero_in_supp}, we can actually identify whether or not $\beta_{0k}=0$ without relying on our conditional moments, since the sign of $\beta_{0k}$ is equal to that of $\E[Y_t - Y_s | X_{-k,s}=X_{-k,t}, X_{k,t}>X_{k,s}]$. The second result on continuous regressors is stronger. It shows that if Assumption \ref{as:cont_regressors} holds, $\beta_0$ is identified up to a scale $c$, with $c$ belonging at most to $(1/\lambda_{T-1}, \lambda_{T-1})$. This directly implies point identification of relative marginal effects. The second result also states that $B$ includes at most $T!-1$ points, and even only 2 points when $T=3$. Importantly, all these result hold for any possible distribution of $\gamma|X$. Thus, point identification may actually hold for many distributions of $\gamma|X$, a point we shall come back to below.

\medskip
The proof of Theorem \ref{thm:main_continuous} relies on the following ideas. In the first case, when $b_k\not\in \{c\beta_{0k}: c \in \{0\}\cup (1/\lambda_{T-1},\lambda_{T-1})\}$, we construct a subset of $\Supp(X)$ of positive probability such that all nonzero $D_j(x;b)$ have the same sign. The result then follows by Lemma \ref{lem:high_level_id}. We use a similar reasoning to prove \eqref{eq:caract_B}. To establish the upper bounds on $|B|$, we exploit the fact that the family of exponential functions $(v\mapsto \exp(\zeta_k v))_{k=1,\dotsc, K}$ with distinct coefficients $\zeta_k$ forms a Chebyshev system \citep[see, e.g.,][Chapter II for the formal definition of such systems]{krein_1977}. This property implies that some key determinants do not vanish, and any non-zero ``exponential polynomial'' $v\mapsto \sum_{k=1}^K \alpha_k \exp(\zeta_k v)$ does not have more than $K-1$ zeros.

\medskip
We now turn to necessary conditions for identification. The following result is a partial converse of Lemma \ref{lem:high_level_id} and Theorem \ref{thm:main_continuous} above.

\begin{thm}\label{thm:nec_conditions}
	Suppose that Assumptions \ref{as:basic}-\ref{as:gen_logit} hold and $T=\tau+1\geq 3$. Then:
	\begin{enumerate}
		\item If $\mathbb{P}\left(|\{X_1, \dotsc, X_T\}| =T\right)=0$, then $B=\R^{K*}$.
		\item If $T=3$, then, for any $b\in R$, there exists a distribution of $\gamma|X$ such that for the corresponding distribution of $Y|X$, $b\in B$.\footnote{Note that $B$ depends on the distribution of $\gamma|X$ but as before, we leave this dependence implicit.}
	\end{enumerate}
\end{thm}

The first result shows that for our conditional moments to have any identifying power, there must exist trajectories of $X=(X_1,\dotsc, X_T)$ with distinct values at all periods. Since we focus here on $T\geq 3$, this excludes in particular the case where $X_t$ is binary. More generally, if all components of $X_t$ are binary, one must have $K> \log(T)/\log(2)$ for our moment conditions to have some identifying power. The second result shows that when $T=3$, one cannot improve \eqref{eq:caract_B}, at least in a uniform sense over conditional distributions of $\gamma$. Specifically, for any $b\in R$, there exists a data generating process satisfying Assumptions \ref{as:basic}-\ref{as:gen_logit} and for which $b\in B$. Note however that failure of point identification at $b$ implies strong restrictions on the distribution of  $\gamma|X$. If $b\in B$ with $b\neq \beta_0$, then, for almost all $x$,
\begin{equation}
E\left[a_1(\gamma, x)D_1(x,b)+ a_2(\gamma, x)D_2(x,b)|X=x\right] =0,	
	\label{eq:mom_non_ident}
\end{equation}
where $a_i(\gamma,x)$ is defined in \eqref{eq:def_a_gamma}. Namely, the distribution of $\gamma|X$ should satisfy a conditional moment restriction (note that \eqref{eq:mom_non_ident} trivially holds when $b$ is replaced by $\beta_0$, because $D_1(x,\beta_0)=D_2(x,\beta_0)=0$). A violation of \eqref{eq:mom_non_ident} on a set of $x$ of positive measure is sufficient to discard $b$ from $B$.\footnote{Related to this, we establish point identification of $\beta_0$ under some restrictions on the conditional distribution of $\gamma|X$  in a \href{https://arxiv.org/abs/2009.08108v1}{previous version} of the paper.}

\section{GMM estimation}\label{sec:estimation}

\subsection{Efficiency bounds} 
\label{sub:estimation_and_efficiency_bounds}

We now suppose point identification based on \eqref{eq:condit_mom} (namely, $B=\{\beta_0\}$) and discuss estimation of $\beta_0$. Let $R(X) = \E[\nabla_{\beta}m(Y,X;\beta_0) | X]$, $\Omega(X) = \V[m(Y,X;\beta_0)|X]$ (so that $\Omega(X) \in \R$) and define, provided that it exists,
$$V_0:=\E \left[ \Omega(X)^{-1}R(X)R(X)'\right]^{-1}.$$
As shown by \cite{chamberlain_1987}, asymptotically optimal estimators of $\beta_0$ based on \eqref{eq:condit_mom} have an asymptotic variance equal to $V_0$. The standard way to construct such estimators consists in two steps: first, one uses the unconditional moment $g(X) m(Y,X;\beta)$ for some $g(\cdot)$ and second, one estimates the optimal instruments $g^\star(X)\triangleq R(X)/\Omega(X)$. Such estimators, however, are  not consistent if
$$\E[g(X) m(Y,X;\beta)]=0 \; \text{ or }\; \E[g^\star(X) m(Y,X;\beta)]=0$$
for $\beta\neq \beta_0$; see \cite{dominguez_lobato_2004}. Instead, we can use an efficient GMM estimator exploiting the continuum of moment conditions associated with \eqref{eq:condit_mom}. We refer in particular to Sections 4 in \cite{HSU201187} and Section 2.5 in \cite{lavergne2013} for the construction of such estimators.

\medskip
These GMM estimators are optimal among those based on \eqref{eq:condit_mom}. However, it is not obvious that \eqref{eq:condit_mom} actually exhausts all the possible restrictions induced by the model, and therefore that $V_0$ is the semiparametric efficiency bound of $\beta_0$. Theorem \ref{thm:semiparam} below shows that this is the case for  $T=\tau+1=3$ under the following conditions.

\begin{hyp}
	\begin{enumerate}
		\item There exists $t \in \{1,...,T\}$ such that $\E[X_tX_t']$ is nonsingular.
		\item  $\E \left[\Omega^{-1}(X) R(X)R(X)'\right]$ exists and is nonsingular.
		\item $|\Supp(\gamma|X)| \geq 10$ almost surely.
	\end{enumerate}
	\label{as:GMM}
\end{hyp}

The first condition is a mild restriction on $X$. The second condition is a local identifiability condition, which is neither weaker nor stronger than $B=\{\beta_0\}$. The third condition is weaker than that imposed by  \cite{chamberlain2010}, namely $\Supp(\gamma|X)=\R$. Intuitively, if $\gamma|X$ has few points of support, moments of $\gamma|X$ are restricted, and we may exploit this to produce additional restrictions that would improve an estimation of $\beta_0$ based solely on \eqref{eq:condit_mom}.

\begin{thm}\label{thm:semiparam}
	Assume $T=\tau+1=3$ with $\lambda_{2} \neq 2$ and Assumptions~\ref{as:basic}, \ref{as:gen_logit} and \ref{as:GMM} hold. Then the semiparametric efficiency bound of $\beta_0$, $V^\star(\beta_0)$, is finite and satisfies $V^\star(\beta_0)=V_0$.
\end{thm}

Intuitively, this result states that all the information content of the model is included in the conditional moment restriction $\E[m(Y,X;\beta_0)|X]=0$. It complements, for  $T=\tau+1=3$, the result of \cite{Hahn1997}, which states that the conditional maximum likelihood estimator is the efficient estimator of $\beta_0$ if $F$ is logistic. Note however that we cannot compare his bound with ours in the logistic case: for this distribution, $w_2=0$, and for identification reasons, this case is excluded from our family of generalized logistic distributions with $\tau=2$. We refer to Footnote \ref{foot:w_zero} above for more detials about this.

\subsection{Unbalanced panel}
\label{sub:unbalanced}

In many applications, as that considered below, panel data are unbalanced. To handle this case, we can simply consider, for each individual, all possible subsets of periods of size $\tau+1$ and form the corresponding moment conditions. Specifically, suppose 
that the set of periods available for individual $i$ is $\mathcal{T}_i \subset \{1,...,T\}$. Thus, we observe the sample $((Y_{it},X_{it})_{t\in \mathcal{T}_i})_{i=1,...,n}$.  Let us assume that the selection of periods is (conditionally) exogenous, namely
\begin{equation}
\mathcal{T}_i \indep (Y_{it})_{t\geq 1} |(X_{it})_{t\geq 1},\gamma_i.	
	\label{eq:exog_sel}
\end{equation}
Then, we basically get back to the case $T=\tau+1$ by considering the moment vector
\begin{align*}
	\psi(Y_i,X_i,\mathcal T_i,\beta)=\ind{|\mathcal{T}_i|\geq \tau+1}\hspace{-0.4cm} \sum_{\substack{t_1< ...<t_{\tau+1} \\ (t_1,...,t_{\tau+1})\in\mathcal{T}_i^{\tau+1}}} \hspace{-0.4cm} g(X_{it_1},...,X_{it_{\tau+1}})m((Y_{it},X_{it})_{t\in\{t_1,...,t_{\tau+1}\}}, \beta).
\end{align*}
for some function $g(X_{it_1},...,X_{it_{\tau+1}}) \in \R^L$, with $L\ge K$. Condition \eqref{eq:exog_sel} ensures that $\E\left[\psi(Y_i,X_i,\mathcal T_i,\beta_0)\right]=0$. Then, we can consider the GMM estimator
\begin{equation}
	\widehat{\beta}=\arg\min_{\beta} \left(\sum_{i=1}^n\psi(Y_i,X_i,\mathcal T_i,\beta)\right)'\widehat{W}\left(\sum_{i=1}^n\psi(Y_i,X_i,\mathcal T_i,\beta)\right), \label{eq:GMM_unbal}
\end{equation}
for some symmetric positive definite $\widehat{W}$. This idea also applies to balanced panel data for which $T>\tau+1$. In such a case, $\mathcal{T}_i=\{1,...,\tau+1\}$ and \eqref{eq:exog_sel} automatically holds.

\section{Application to \cite{BrenderDrazen2008}}
\label{sec:appli}

\cite{BrenderDrazen2008} study how budget deficits and economic growth affect reelection. To this end, they gather data from multiple sources on 74 countries, over the period 1960-2003. They use two definitions for their binary outcome variable REELECT, one where reelection is defined in a ``narrow'' sense and another where it is ``expanded'', following here their terminology. This also leads to two different samples, as REELECT may be missing in the narrow sense but equal to 0 in the expanded sense. The covariates related to budget deficits are BALCH\_term and BALCH\_ey. BALCH\_term corresponds to the change in ratio of the central government's balance to GDP over the term in office. BALCH\_ey is the change in the balance/GDP ratio between the year preceding the election and the election year. The variable GDPPC\_gr is the average annual growth rate of real GDP per capita between two election years. The authors also include in their models two controls, namely a dummy for a new democracy and a dummy of having a majoritarian electoral system. We refer to \cite{BrenderDrazen2008} for more details about the data.

\medskip
In their main specification, \cite{BrenderDrazen2008} consider a simple logit model, see Table 2 therein. Then, as a robustness check (see their Table 3), they estimate a fixed effect logit model. They show that their main results are robust to including fixed effects. However, the assumption of logistic errors is not warranted, so we investigate whether the results are robust to this assumption, by considering instead the family of generalized logistic distribution, with $\tau=2$. 
We focus on the sample of developed countries as the sample of less developped countries is very small, and thus leads to noisy estimates. Note that the data are not balanced at all: some countries are only observed for $4$ periods in the narrow sample (resp.~$5$ in the expanded sample), while others are observed over $13$ (resp.~$14$) periods. We thus apply the procedure mentioned in Section \ref{sub:unbalanced}. The vector of instruments $g(X_{it_1},X_{it_2},X_{it_3})$ is simply the list of the corresponding 15 variables (as $X_{it}\in\R^5$), demeaned over these three periods. We consider $\lambda_2 = 1.2, 1.4, 1.6$ and $1.8$. We do not consider larger values of $\lambda_2$ as they seem to lead to numerical instabilities.\footnote{This may be because $|M_t(x;\beta)|$ increases quickly with $\lambda_2$, due to the exponential function.} Finally, as the GMM objective function may have local optima, we consider 200 random initial points and pick the vector of parameters minimizing the corresponding final objective function.

\medskip
The results are presented in Table \ref{tab:results_appli}. Because the coefficients themselves are not comparable, we focus on the sign of BALCH\_ey and on the relative effects with respect to BALCH\_ey; note that we were able to recover the exact same estimates as \cite{BrenderDrazen2008} in their Tables 2 and 3. We choose BALCH\_ey as the reference variable for relative effects because its coefficient should not be 0, and it has the largest t-test on the logit and fixed effect logit model. For the three methods, the $t$-statistics of relative effects under the null hypothesis are obtained using the estimated asymptotic variance of $\widehat{\beta}$.

\medskip
Overall, at least two important results seem robut to the distributional assumption on the unobserved terms. First, the sign of BALCH\_ey is always positive. Second, the relative effect of BALCH\_term and BALCH\_ey remain quite stable when considering our FE generalized logistic model, with fluctuations between 0.27 and 0.52 depending on the sample and value of $\lambda_2$ that we consider. At the 10\% level, we cannot reject that the effect of  BALCH\_term is actually 0, except in the narrow sample with $\lambda_2=1.8$. But the test was already close to not being rejected with the FE logit model on the narrow sample (p-value=0.097), and not rejected with the simple and FE logit models based on  the expanded sample (p-values=0.124 and 0.204 respectively). So the most important results seem overall robust to the change of specification we consider. Other results fluctuate slightly more: the fact of being a new democracy had a positive and borderline significant effect with the expanded sample (p-value=0.099). It is not significant anymore with our model, the coefficient being  sometimes even negative.

\begin{table}[H]
	\caption{Estimates of relative effects of budget balances and growth on the probability of reelection in developed economies}
	\label{tab:results_appli}
	\centering	
	\begin{adjustbox}{max width={0.99\linewidth},center}
		\begin{threeparttable}[h]
			\begin{tabular}{lcccccc}
				\toprule
				{} &      Logit &   FE logit &  \multicolumn{4}{c}{FE generalized logit } \\
				\cline{4-7}
				$\lambda_2$ &      {} &   {} &  1.2 & 1.4 & 1.6 & 1.8 \\
				\midrule
				{\bf Narrow sample} \\
				Sign of BALCH\_ey & >0 & >0 & >0 & >0 & >0 & >0 \\
				BALCH\_term/BALCH\_ey  &   0.54 &   0.55 &  0.48 &   0.37 &   0.37 &   0.52 \\
				&   (2.34) &   (1.82) &  (0.03) &   (0.06) &   (0.03) &   (2.49) \\
				
				GDPPC\_gr/BALCH\_ey  &  -0.04 &   0.27 &  -0.35 &  -0.34 &  -0.33 &  -0.37  \\
				&   (0.17) &  (0.70) &  (0.04) &   (0.07) &   (0.04) &   (1.04) \\
							
				New democraties/BALCH\_ey  &    0.03 &   0.07 &   0.05 &   0.05 &   0.05 &  -0.20 \\
				& (2.69) &   (1.62) &  (0.04) &  (0.08) &  (0.04) &   (0.03) \\
				
				Majoritarian   &   0.02 &   0.07 &    -0.10 &  0.00 &  0.00 &  0.03 \\
				electoral system/BALCH\_ey  &  (1.31) &   (1.52) &   (0.03) &   (0.02) &   (0.01) &   (0.20) \\
			
				\midrule
				{\bf Expanded sample} \\
				Sign of BALCH\_ey & >0 & >0 & >0 & >0 & >0 & >0 \\
				BALCH\_term/BALCH\_ey  &   0.40 &   0.36 &  0.34 &   0.27 &   0.37 &   0.47 \\
				&   (1.44) &   (1.35) &   (0.85) &   (1.64) &   (0.72) &   (0.42)   \\
				
				GDPPC\_gr/BALCH\_ey &  0.09 &   0.46 &  -0.09 &   0.00 &  -0.14 &  -0.28 \\
				&  (0.30) &   (1.20) &   (0.22) &   (0.00) &   (0.29) &   (0.61) \\
				
				New democraties/BALCH\_ey &   0.04 &   0.09 &    0.02 &   0.01 &  -0.00 &  -0.15 \\
				&    (3.11) &   (1.81) &   (0.32) &   (0.08) &   (0.00) &   (0.26) \\
				
				Majoritarian &  0.02 &   0.04 &  -0.15 &  -0.10 &  -0.21 &  -0.67 \\
				electoral system/BALCH\_ey  &   (1.74) &   (1.12) &   (0.99) &   (1.54) &   (1.08) &   (1.14) \\
				\bottomrule
			\end{tabular}
			\begin{tablenotes}
				\footnotesize \item {\em Notes:} Analytical t-statistics of the coefficient ratios are under parentheses. The estimated asymptotic variance of the simple logit model is obtained through clustering at the country level. Both samples include 22 countries, with on average $7.1$ (resp.~$7.9$) periods per country in the narrow (resp.~expanded) sample.
			\end{tablenotes}
		\end{threeparttable}
	\end{adjustbox}
\end{table}

\section{Conclusion}\label{sec:conclusion}
	
This paper studies the identification and root-n estimation of the common slope parameter in a static panel binary model with exogenous and bounded regressors. We first show that when $T\geq 3$ and the unobserved terms belong to a family of generalized logistic distribution, a conditional moment restriction holds. Then, we study the identified set corresponding to these restrictions. In particular, under a restriction on the distribution of covariates only, relative effects are point identified, no matter the distribution of the individual effect. Our identification results lead to a GMM estimator that reaches the semiparametric efficiency bound when $T=3$. Estimating this model may serve as a robustness check for the fixed effect logit model, something we illustrate in the application.

\medskip
Our paper also leaves a few questions unanswered. A first one is whether the family of $F$ considered here is the only one for which point identification can be achieved. Another one is whether the GMM estimator still reaches the semiparametric efficiency bound when $T>3$. Both questions raise difficult issues and deserve future investigation.
	
\newpage
\bibliography{../biblio}
	
\newpage
\appendix

\section{Proofs of the results}\label{sec:proofs}
	
For any real $a \in \R$, we let $\text{sgn}(a) := \mathds{1}(a> 0) - \mathds{1}(a< 0)$. For any subset $A$ of a reference space $E$, we let $A^c$ denote the complement of $A$ in $E$. The following lemma on ``exponential polynomials'' is key in the proof of Theorems~\ref{thm:main_continuous} and \ref{thm:semiparam}.
	
\begin{lem}\label{lem:pol_exp}
Let $K\geq 1$, $(\zeta_1, \dotsc, \zeta_K)$ be $K$ distinct real numbers, $(\alpha_1,\dotsc,\alpha_K)'\in\R^K$, $(\alpha_1,\dotsc,\alpha_K)\neq (0,\dotsc,0)$ and $P(x):=\sum_{k=1}^K \alpha_k \exp(\zeta_k x)$. Then $P$ has at most $K-1$ distinct roots.
\end{lem}

The proof is by induction on $K$ and Rolle's theorem, see e.g. Chapter 2, section 2 of \cite{krein_1977}.

\subsection{Proposition~\ref{prop:beta0_0}}\label{sec:proof_prop_beta0_0}
	
The sufficient part is obvious. To prove necessity, suppose $\beta_0\neq 0$. Since $\E[(X_t-X_{t'})(X_t-X_{t'})']$ is non singular, there exists a subset $\mathcal{S}$ of the support of $(X_t,X_{t'})$ such that $\P(\mathcal{S})>0$ and for all $(x_t,x_{t'})\in\mathcal{S}$, $(x_t-x_{t'})'\beta_0$ has constant, non-zero sign. Without loss of generality let us assume $(x_t-x_{t'})'\beta_0>0$. Let $G(u):=F(u)/(1-F(u))$. Because $G$ is strictly increasing, we have, for all $g\in\R$,
$$G(x_t'\beta_0+g)>G(x_{t'}'\beta_0+g).$$
Equivalently,
$$F(x_t'\beta_0+g)(1-F(x_{t'}'\beta_0+g))>F(x_{t'}'\beta_0+g)(1-F(x_t'\beta_0+g)).$$
In other words,
$$\P(Y_1=1,Y_{t'}=0|X_t=x_t,X_{t'}=x_t', \gamma=g)>\P(Y_1=0,Y_{t'}=1|X_t=x_t,X_{t'}=x_t', \gamma=g),$$
and the result follows by integration over $g$.
	
\subsection{Theorem~\ref{thm:moments}}
	
Let us define
$$ A(x, \gamma; \beta) \triangleq
\begin{pmatrix}
\sum_{j=1}^{T-1} w_j \exp(\lambda_{j} (x_1'\beta+\gamma)) & \ldots & \sum_{j=1}^{T-1} w_j \exp(\lambda_{j} (x_T'\beta+\gamma))\\
\exp(\lambda_{1}x_1'\beta) & \ldots & \exp(\lambda_{1}x_T'\beta) \\
\vdots &  & \vdots \\
\exp(\lambda_{T-1}x_1'\beta) & \ldots & \exp(\lambda_{T-1}x_T'\beta)
\end{pmatrix}.$$
Let $A_i(x, \gamma; \beta) $ denote the $i$th row of $A(x, \gamma; \beta)$. Then
$$A_1(x, \gamma; \beta)=\sum_{j=1}^{T-1} w_j \exp(\lambda_{j} \gamma)A_{j+1}(x, \gamma; \beta).$$
It follows that for all $(x,\gamma) \in \Supp(X) \times \R$,
$$\det A(x, \gamma;\beta_0)= 0.$$
By Assumption~\ref{as:gen_logit} and since we focus on the first type therein, we have $G(u) \triangleq F(u)/(1-F(u)) = \sum_{j=1}^{T-1}w_j\exp(\lambda_{j}u)$. Now, developing $\det A(x,\gamma; \beta_0)$ with respect to the first row yields, by definition of the function $m$,
$$\sum_{y\in\{0,1\}^T}m(y,x; \beta_0)\prod_{t:y_t=1}G(x_t'\beta_0 + \gamma) = 0.$$
Multiplying this equality by $\prod_t(1-F(x_t'\beta_0 + \gamma))$ we obtain
$$\sum_{y\in\{0,1\}^T} \left[m\left(y,x; \beta_0 \right) \prod_{t:y_t=1}F(x_t'\beta_0 + \gamma)\prod_{t:y_t=0}(1-F(x_t'\beta_0 + \gamma))\right] = 0.	$$
This equation is equivalent to $\E\left [m(Y,X;\beta_0) | X, \gamma \right] = 0$ a.s. The result follows.

\subsection{Lemma \ref{lem:high_level_id}}

Let $b\in \widetilde{B}^c$ and let us prove that $b\not\in B$. Fix $x \in \mathcal{D}(b)$ and let $\mathcal J_x:=\{j \in \{1,...,T-1\} :  D_j(x;b) \neq 0\}$ and
\begin{equation}
a_j(x) := w_j\E\left[\frac{\exp(\lambda_{j}\gamma)}{\prod_{t=1}^T\left(1 + \sum_{k=1}^{T-1} w_k \exp(\lambda_{k}(x_t'\beta_0 + \gamma)) \right)} \Big |X=x \right].	
	\label{eq:def_a}
\end{equation}
Then $\mathcal J_x \neq \emptyset$ and
\begin{equation}\label{eq:moment_fail_lin_case}
	\E[m(Y,X;b)|X=x] = \sum_{j\in\mathcal J_x} a_j(x)D_j(x;b).
\end{equation}
Moreover, $a_j(x)>0$ and all the $D_j(x;b)$ for $j \in \mathcal J_x$ have the same sign. Thus,  $\E[m(Y,X;b)|X=x]\neq 0$. Because $b\in \widetilde{B}^c$, we have, by definition of $\widetilde{B}$, $\P(X \in \mathcal{D}(b))>0$. Thus, $\E[m(Y,X;b)|X=x]\neq 0$ with positive probability, implying  $b\not\in B$.

\subsection{Theorem \ref{thm:main_continuous}}
\subsubsection*{Part 1} 

\noindent {\bf a. $B_k\subset R_k:=\{c\beta_{0k}: c\in \{0\}\cup(1/\lambda_{T-1},\lambda_{T-1})\}$.}

\medskip
Let us fix $k\in\{1,...,K\}$, $b=(b_1,...,b_K)$ and define
$$\mathcal{X}_{0k}:=\left\{x\in \Supp(X): \, x_{j,1}=...=x_{j,T}=0 \; \forall j\neq k, |\{x_{k,1},...,x_{k,T}\}|=T\right\}.$$
First, suppose that $\beta_{0k}=0$ and $b_k\neq 0$. Then, $D_j(x;b)$ does not depend on $j$. Moreover, because $$|\{x_1'b,...,x_T'b\}| = |\{x_{k,1}b_k,...,x_{k,T}b_k\}|=T,$$
we have $D_j(x;b)\neq 0$ by properties of Chebyshev systems. Thus, $x\in \mathcal{D}(b)$, implying that $\mathcal{X}_{0k}\subset \mathcal{D}(b)$. By Assumption \ref{as:zero_in_supp}, $\P(X\in \mathcal{X}_{0k})>0$. Hence, $\P(X\in \mathcal{D}(b))>0$. By Lemma \ref{lem:high_level_id}, $b_k \not\in B_k$ and $B_k\subset\{0\}=R_k$.

\medskip
Now, suppose $\beta_{0k}\neq 0$. Then any $b_k\in \R$ can be written as $c\beta_{0k}$. We prove that if $c \not\in\{0\}\cup (1/\lambda_{T-1},\lambda_{T-1})$, then $\mathcal{X}_{0k}\subset \mathcal{D}(b)$. By Lemma \ref{lem:high_level_id} again, this shows that $B_k\subset R_k$. Let us first suppose that $c\not\in\{1/\lambda_{T-1},\lambda_{T-1}\}$ and fix $x\in\mathcal{X}_{0k}$. Let us show that for each ($j,j')\in \{1,...,T-1\}^2$,
\begin{equation}\label{eq:same_sign}
\mathrm{sign}(D_j(x;b))=\mathrm{sign}(D_{j'}(x;b)) \neq 0.
\end{equation}
If $c \in (-\infty, 0)$ , we have
\begin{equation}\label{eq:case_includ_left}
c\lambda_{T-1}<...< c\lambda_{2}< c < 0<1<\lambda_2 < ...< \lambda_{T-1}.
\end{equation}
If $c \in (0, 1/\lambda_{T-1})$, we have
\begin{equation}\label{eq:case_includ_left2}
0< c < c\lambda_{2}< ...<c\lambda_{T-1}<1<\lambda_2<  ...< \lambda_{T-1}.
\end{equation}
Else, $c \in (\lambda_{T-1},+\infty)$ and we have
\begin{equation}\label{eq:case_includ_left3}
1<\lambda_{1} < ...< \lambda_{T-1} < c < c\lambda_{2}< ...<	c\lambda_{T-1}.
\end{equation}
Let $p_j$ denote the number of transpositions (ie permutations exchanging two elements, leaving the others fixed) needed to sort $\tilde \lambda^j:=(\lambda_{j}, c ,c\lambda_{2}, ..., c\lambda_{T-1})'$ in ascending order. It is clear from Equations \eqref{eq:case_includ_left}-\eqref{eq:case_includ_left3} that $p_j=p_{j'} = p$ for all $(j,j')\in\{1,...,T-1\}^2$. Let $\tilde \lambda^{sj}$ denote the sorted version of $\tilde \lambda^j$ and define
\begin{equation*}
D_j(x; b, \lambda) =: \det \begin{pmatrix} \exp(\lambda_jx_1'\beta_0) & \ldots & \exp(\lambda_jx_T'\beta_0) \\
\exp(\lambda_1x_1'b) & \ldots & \exp(\lambda_1x_T'b) \\
\vdots & & \\
\exp(\lambda_{T-1} x_1'b) & \ldots & \exp(\lambda_{T-1} x_T'b)
\end{pmatrix},
\end{equation*}
so that $D_j(x; b) = D_j(x;b, \lambda)$. Because $x\in \mathcal{X}_{0k}$, we have
$$D_j(x; b, \lambda) = \det \begin{pmatrix} \exp(\lambda_j x_{k,1}\beta_{0k}) & \ldots & \exp(\lambda_jx_{k,T}\beta_{0k}) \\
\exp(c \lambda_1 x_{k,1}\beta_{0k}) & \ldots & \exp(c \lambda_1 x_{k,T}\beta_{0k}) \\
\vdots & & \\
\exp(c\lambda_{T-1} x_{k,1}\beta_{0k}) & \ldots & \exp(c\lambda_{T-1} x_{k,T}\beta_{0k})
\end{pmatrix} = D_j(x ; \beta_0, \tilde \lambda^j).$$
Hence, for all $j\in\{1,...,T-1\}$,
\begin{align*}
\text{sgn}\left(D_j(x; b) \right) & = \text{sgn}\left(D_j(x ; \beta_0, \tilde \lambda^j)\right) = (-1)^{p}\text{sgn}\left(D_j(x; \beta_0, \tilde \lambda^{sj})\right).
\end{align*}
Now, let $\overline{p}$ be the number of pairwise coordinates permutations needed to sort the vector $(x_1'\beta_0, ..., x_T'\beta_0)'$ in ascending order, and let $x^s$ denote a rearrangement of $x$  such that $x^s_1{}'\beta_0 < ...< x^s_T{}'\beta_0$. It follows that, for all $j\in\{1,...,T-1\}$,
\begin{align*}
\text{sgn}\left(D_j(x; b) \right) & = 	(-1)^{p}\text{sgn}\left(D_j(x ; \beta_0, \tilde \lambda^{sj})\right) \\
& = (-1)^{p + \overline{p}}\text{sgn}\left(D_j(x^s; \beta_0,  \tilde \lambda^{sj})\right) \\
& = (-1)^{p + \overline{p}},
\end{align*}
where the last equality follows by properties of Chebyshev systems. The last equality implies that  \eqref{eq:same_sign} holds. Hence $x\in \mathcal{D}(b)$, implying $\P(X \in \mathcal{D}(b))>0$.

\medskip
Finally, consider the case where $b=c\beta_0$ with $c\in\{1/\lambda_{T-1},\lambda_{T-1}\}$. By continuity of the determinant and \eqref{eq:same_sign}, we either have $0\leq \min_{j=1,...,T-1} D_j(x;b) \leq  \max_{j=1,...,T-1} D_j(x;b)$ or  $0\geq \max_{j=1,...,T-1} D_j(x;b) \geq  \min_{j=1,...,T-1} D_j(x;b)$. Moreover, $D_{T-1}(x;\beta_0/\lambda_{T-1}) \neq 0$ and $D_1(x;\lambda_{T-1}\beta_0) \neq 0$. Therefore, whatever the value of $c$ ($1/\lambda_{T-1}$ or $\lambda_{T-1}$), we have $x\in \mathcal{D}(b)$. Then, again, $\P(X \in \mathcal{D}(b))>0$. The result follows.

\medskip
\noindent{\bf b. $|B|<\infty$.}

\medskip
Because $|B|\leq \prod_{k=1}^K |B_k|$, it suffices to prove that for each $k$, $|B_k|<\infty$. Fix $k$. If $\beta_{0k}=0$, then $B_k=\{0\}$ and we have nothing to prove. Otherwise, let $b=(b_1,...,b_K)\in B$ and fix $x=(x_1,...,x_T)\in\mathcal{X}_{0k}$. Let $c\in \{0\}\cup (1/\lambda_{T-1},\lambda_{T-1})$ be such that $b_k=c\beta_{0k}$. By Equation \eqref{eq:moment_fail_lin_case}, we have $\sum_{j=1}^{T-1} a_j(x) D_j(x; b)=0$, where $a_j(x)$ is defined by \eqref{eq:def_a}. Moreover, by definition of $\mathcal{X}_{0k}$, we have $D_j(x; b)=D_j(x;c\beta_0)$. Then, $c$ satisfies
\begin{equation}\label{eq:moment_lin_case}
\sum_{j=1}^{T-1} a_j(x) D_j(x; c\beta_0)=0,
\end{equation}
Developing $ D_j(x; c\beta_0)$ with respect to the first line, and using the definition of the determinant, we obtain
\begin{equation}
\sum_{t=1}^T (-1)^{t+1} \sum_{j=1}^{T-1} a_j(x) \exp(\lambda_{j} x_t'\beta_0) \sum_{\sigma \in \mathfrak{S}_t} \eps(\sigma) \exp\left[\left(\sum_{s\neq t} \lambda_{\sigma(s)} x_{s}'\beta_0\right)c\right] =0,	
\label{eq:c}
\end{equation}
where $\mathfrak{S}_t$ is the set of bijections from $\{1,...,T\}\backslash\{t\}$ to $\{1,...,T-1\}$ and $\eps(\sigma)$ denotes the parity of $\sigma$ (we can assimilate $\sigma$ to a permutation by assimilating $\{1,...,T\}\backslash\{t\}$ with $\{1,...,T-1\}$, keeping the natural ordering of both sets). The left-hand side of \eqref{eq:c} is a function of $c$ of the form $\sum_{k=1}^{K} d_k \exp(b_k c)$, with $K\leq T!$ (the inequality arises because some coefficients in the exponential monomials may be equal). Let us show that $d_k\neq 0$ for at least one $k$. First, remark that $x_t'\beta_0 = x_{k,t}\beta_{0,k}$. Then, because $|\{x_{k,1},...,x_{k,T}\}|=T$ and $\beta_{0k}\neq 0$, we can assume without loss of generality, up to a rearrangement of periods, that $x_1'\beta_0 <...<x_T'\beta_0$. Let $I_t$ be the element of $\mathfrak{S}_t$ such that $I_t(s)=s-\mathds{1}_{\{s\geq t+1\}}$. By the rearrangement inequality, for all $\sigma\in\mathfrak{S}_t\backslash\{I_t\}$,
$$\sum_{s\neq t} \lambda_{\sigma(s)} x_{s}'\beta_0< \sum_{s\neq t} \lambda_{I_t(s)} x_{s}'\beta_0.$$
Moreover, for all $t\in\{1,...,T\}$, let
$$g(t):=\sum_{s\neq t} \lambda_{I_t(s)} x_{s}'\beta_0.$$
Because $I_t(s)=I_{t-1}(s)$ for all $t>1$ and $s\leq t-2$ or $s\geq t+1$, we have
$$g(t)-g(t-1)=\lambda_{t-1}x_{t-1}'\beta_0-\lambda_{t-1}x_t'\beta_0 < 0.$$
Hence, the exponential monomial with highest coefficient in \eqref{eq:c} is
$$\exp\left[\left(\sum_{s\neq 1} \lambda_{s-1} x_{s}'\beta_0\right)c\right]$$
and we can obtain it only by letting  $t=1$ and $\sigma=I_1$. Because $a_j(x)>0$ for all $j$, the  coefficient of this monomial is $\sum_{j=1}^{T-1} a_j(x) \exp(\lambda_{j} x_1'\beta_0)>0$. Therefore, at least one $d_k$ in the exponential polynomial $\sum_{k=1}^K d_k \exp(b_k c)$ satisfies $d_k\neq 0$. Then, by Lemma \ref{lem:pol_exp}, the equation $\sum_{k=1}^K d_k \exp(b_k c)=0$ has at most $T!-1$ solutions. Thus, $|B_k|\leq T!-1$. The result follows.

\subsubsection*{Part 2}
\label{ssub:_eqref_holds}

The point identification of relative marginal effects is obvious given the other results, which we prove in turn.

\medskip
\noindent\textbf{a. Equation \eqref{eq:caract_B} holds.}

\medskip
Let us define, for all $b \in \R^{K*}$,
\begin{align*}
\mathcal{X}_1(b) & = \left\{x=(x_1,...,x_T)\in \text{Supp}(X): \exists (s,t)\in\{1,...,T\}^2: x_s'b = x_t'b, \; x_s'\beta_0\neq x_t'\beta_0, \right.
\\
& \left. \qquad \text{and } \forall (s',t') \in \{1,...,T\}^2, s'\neq t', \{s',t'\} \neq \{s,t\}: \; x_{s'}'b\neq x_{t'}'b\right\}.
\end{align*}
The proof of is divided into three steps. First, we prove that $\mathcal X_1(b)\subset \mathcal{D}(b)$, for all $b\in\R^{K*}\backslash \lin(\beta_0)$. In a second step, we prove that $\widetilde{B}\subset \lin(\beta_0)$. Finally, the third step shows that $\widetilde{B}\subset R$.

\medskip
\noindent {\it First step: $\mathcal X_1(b)\subset \mathcal{D}(b)$ for all $b \in \R^{K*}\backslash\lin(\beta_0)$.}

\medskip
Let $x \in \mathcal X_1(b)$ and $(s,t)$ be as in the definition of $\mathcal X_1(b)$. Developing $D_j(x;b)$ according to the first row, we obtain, for all $j\in\{1,...,T-1\}$,
$$D_j(x;b) = \sum_{\ell=1}^{T}(-1)^{\ell+1}\exp(\lambda_{j}x_\ell'\beta_0)D_j^{-\{1,\ell\}}(x;b),$$
where $D_j^{-\{1,\ell\}}(x;b)$ denotes the determinant of the matrix in  $D_j(x;b)$ once its first row and $\ell$th column have been removed. Remark that, for all $j\in\{1,..., T-1\}$, for all $\ell \in\{1,...,T\}\backslash\{s,t\}, \; D_j^{-\{1,\ell\}}(x;b) = 0$, and
$$D_j^{-\{1,t\}}(x;b) = (-1)^{\vert s - t\vert  - 1} D_j^{-\{1,s\}}(x;b).$$
As a result,
\begin{align*}
D_j(x;b) & = (-1)^{s+1}\exp(\lambda_{j}x_s'\beta_0)D_j^{-\{1,s\}}(x;b) + (-1)^{t+1}\exp(\lambda_{j}x_t'\beta_0)D_j^{-\{1,t\}}(x;b) \\
& = D_j^{-\{1,s\}}(x;b) \left[(-1)^{s+1}\exp(\lambda_{j}x_s'\beta_0) + (-1)^{t+1}\exp(\lambda_{j}x_t'\beta_0)(-1)^{|s-t|-1}\right] \\
& = D_j^{-\{1,s\}}(x;b) (-1)^{s+1}\left[\exp(\lambda_{j}x_s'\beta_0) - \exp(\lambda_{j}x_t'\beta_0) \right],
\end{align*}
where we have used $(-1)^{|s-t|+t}=(-1)^s$. Now, $D_j^{-\{1,s\}}(x;b)$ does not depend on $j$ and by definition of Chebyshev systems, $D_j^{-\{1,s\}}(x;b)\neq 0$. Also, the sign of the term inside brackets is equal to the sign of $(x_s-x_t)'\beta_0$, and thus does not depend on $j$. Hence for all $(j,j')\in\{1,...,T-1\}^2$,
$$\mathrm{sgn}\left(D_j(x;b)\right)=\mathrm{sgn}\left(D_{j'}(x;b)\right) \neq 0,$$
which shows that $x \in \mathcal{D}(b)$.

\medskip
\noindent{\it Second step: $\widetilde{B}\subset \lin(\beta_0)$.}

\medskip
Fix $b\not\in \lin(\beta_0)$, $b\neq 0$ and let us prove that $\P(X\in\mathcal{D}(b))>0$. The result will then follow by Lemma \ref{lem:high_level_id}.

\medskip
Suppose without loss of generality that $(s,t)$ in Assumption \ref{as:cont_regressors} is equal to $(1,2)$. By that  assumption, there exists $\widetilde{x}:=(x',x',x_3'...,x_T')'\in\Supp(X)$ and a neighborhood $\widetilde{V}$ of $\widetilde{x}$ included in $\Supp(X)$. Since $b$ and $\beta_0$ are not collinear, there exists $(u'_1,u'_2)'\in\R^{2K}$ such that $(u_1-u_2)'b= 0$ and $(u_1-u_2)'\beta_0 \neq 0$. Moreover, up to replacing $(u'_1,u'_2)'$ by $c(u'_1,u'_2)'$ with $c\neq 0$, $(u'_1,u'_2)'$ can be chosen of arbitrarily small norm.

\medskip
Now, let $x_1=x_2=x$ and
\begin{align*}
\mathcal{A}(u_1,u_2) & = \left\{(u'_3,...,u'_T)'\in \R^{K(T-2)}: \forall (s,t) \in \{1,...,T\}^2, s\neq t, \{s,t\} \neq \{1,2\}: \right. \\ &
\qquad \left. (u_s -u_t + x_s - x_t)'b  \neq 0\right\},
\end{align*}
The set $\mathcal{A}(u_1,u_2)$ is dense as the intersection of open, dense subsets of $\R^{K(T-2)}$. Hence, there exists $(u'_3,...,u'_T)'\in \mathcal{A}(u_1,u_2)$ with arbitrarily small norm. Then, we can ensure that $u:=(u'_1,...,u'_T)'$ satisfies $x^*:=\widetilde{x}+u\in\widetilde{V}$. Moreover, by construction, $x^*\in \mathcal{X}_1(b)$. Then, Step 1 implies $x^* \in \mathcal D(b)$ and $D_j(x;b)\neq 0$ for all $j$. By continuity of the map $x\mapsto D_j(x;b)$ and Assumption \ref{as:cont_regressors}, there exists a neighborhood of $x^*$, $\mathcal V\subset\mathcal D(b)$ such that $\P(X \in \mathcal V)>0$. Hence, $P(X \in \mathcal{D}(b))>0$.

\medskip
\noindent{\it Third step: $\widetilde{B}\subset R$.}

\medskip
We just have to prove that if $b = c\beta_0$ with $c\in(-\infty, 1/\lambda_{T-1}]\cup [\lambda_{T-1},+\infty)$ and $c\neq 0$ (since $\beta_0\neq 0$), then $b\not \in B$. The reasoning is exactly the same as in Part 1.a, with just one change: Instead of considering $x\in\mathcal{X}_{0k}$, we consider $x\in\mathcal{X}_0$, with
$$\mathcal{X}_0:=\left\{x\in \Supp(X): \, |\{x_1'\beta_0,...,x_T'\beta_0\}|=T\right\}.$$

\medskip
\noindent{\bf b. $|B|\leq T!-1$.}

\medskip
The reasoning is exactly the same as in Part 1.b, with just two changes. First, we reason directly on $B$,  not on $B_k$. Second, instead of considering $x\in\mathcal{X}_{0k}$, we consider $x\in\mathcal{X}_0$.

\medskip
\noindent {\bf c. $|B|\leq 2$ when  $T=3.$}

\medskip
For any $b=c\beta_0\in B$, we have, as in Eq. \eqref{eq:moment_lin_case},
\begin{equation}\label{eq:subcase_cond}
a_1(x) D_1(x;c\beta_0) + a_2(x) D_2(x;c\beta_0) = 0
\end{equation}
for almost all $x \in \Supp(X)$. Suppose there exist three distinct solutions $1, c_1, c_2$ to Equation \eqref{eq:subcase_cond}, with $1/\lambda_{2}< c_1<c_2< \lambda_{2}$. Multiply Eq. \eqref{eq:subcase_cond}, evaluated at $c=c_1$, by $D_2(x;c_2\beta_0)$. Similarly, multiply Eq. \eqref{eq:subcase_cond}, evaluated at $c=c_2$, by $D_2(x;c_1\beta_0)$. Substracting the two expressions, we obtain, since $a_1(x)>0$,
\begin{equation}
D_1(x;c_1\beta_0) D_2(x;c_2\beta_0) -D_1(x;c_2\beta_0) D_2(x;c_1\beta_0) =0.	
\label{eq:long_cheby_syst}
\end{equation}
For any $x\in \mathcal{X}_0$, let $u_t:=x_t'\beta_0$. Fixing $u_2$ and $u_3$, \eqref{eq:long_cheby_syst} may be written as
\begin{equation}
P(u_1):=\sum_{k=1}^{13} \alpha_k \exp(\zeta_k u_1) =0,	
\label{eq:poly_u1_zero}
\end{equation}
where the $\alpha_k$ and $\zeta_k$ are functions of $(u_2,u_3)$. Suppose first that $c_2>1$.  Some tedious algebra shows that the smallest $\zeta_k$ is $1+c_1$, and its associated coefficient is equal to
\begin{align*}
\alpha_k = & \left[\exp(c_2 (u_2 + \lambda_{2} u_3)) - \exp(c_2 (u_3 + \lambda_{2} u_2))\right] \\
\times & \left[
\exp(\lambda_{2} (u_2 + c_1 u_3)) - \exp(\lambda_{2} (u_3 + c_1 u_2))\right].	
\end{align*}
Because $u_2\neq u_3$ (as $x\in\mathcal{X}_0$), $\alpha_k \neq 0$. Hence $P$ is nonzero and by Lemma \ref{lem:pol_exp}, it has at most 12 zeros. However, under Assumption \ref{as:cont_regressors}.2 and the second part of Assumption \ref{as:cont_regressors}.3,
$$\left|\Supp(X_1'\beta_0|X_2'\beta_0=u_2,X_3'\beta_0=u_3)\backslash\{u_2,u_3\}\right|>12.$$
Thus, in view of \eqref{eq:poly_u1_zero}, $P$ has strictly more than 12 zeros, a contradiction.

\medskip
Second, suppose that $c_2< 1$. Then, the largest $\zeta_k$ is $\lambda_{2}(1+c_2)$, and its associated coefficient is equal to
\begin{align*}
\alpha_k = -& \left[\exp(u_2 + c_2 u_3)) - \exp(u_3 + c_2 u_2)\right]\\
\times & \left[
\exp(c_1 (u_2 + \lambda_{2} u_3)) - \exp(c_1 (u_3 + \lambda_{2} u_2))\right].
\end{align*}
Again, $\alpha_k\neq 0$ and we reach a contradiction as before. The result follows.

\subsection{Theorem \ref{thm:nec_conditions}} 
\label{sub:thm_nec_cond}

\subsubsection*{Part 1}

Let us suppose that $\mathbb{P}\left(|\{X_1, \dotsc, X_T\}| =T\right)=0$. Let $T_1$ and $T_2>T_1$ denote the two random dates, functions of $X$ only, such that $X_{T_1} = X_{T_2}$ almost surely. For all $t\in\{1,...T\}$, let $e_t$ denote the vector of $T-1$ zeros and a 1 at coordinate $t$. Let  $f(x; b) \triangleq \E\left [m(Y,X;b) | X=x\right]$. By definition,
	\begin{equation}
	f(X;b) = \sum_{y\in\{0,1\}^T} \P(Y=y|X) m(y,X;b).	
	\label{eq:link_f_mu}
	\end{equation}
	Moreover,  almost surely,
	\begin{align}
 \P(Y=e_{T_1}|X) = & \int F(X_{T_1}'\beta_0 + \gamma) \prod_{t\neq T_1} (1 - F(X_t'\beta_0 + \gamma))\mathrm{d}F_{\gamma|X}(\gamma) \nonumber\\
	= & \int F(X_{T_2}'\beta_0 + \gamma) \prod_{t\neq T_2} (1 - F(X_t'\beta_0 + \gamma))\mathrm{d}F_{\gamma|X}(\gamma) \nonumber\\
	=&  \P(Y=e_{T_2}|X). \label{eq:for_f1}
	\end{align}
	Next, remark that the matrices in $M_{T_1}(X;b)$ and $M_{T_2}(X;b)$ have the same columns but in different order, with $T_2 - T_1-1$ transpositions needed to obtain the same ordering. Thus, by definition of the determinant, $M_{T_1}(X;b)= - M_{T_2}(X;b)$, which implies
\begin{equation}
m(e_{T_1},X;b) = - 	m(e_{T_2},X;b).
	\label{eq:for_f2}
\end{equation}
Moreover, for all $s\not\in\{T_1,T_2\}$, $m(e_s,X;b)=0$ because $M_s$ includes two identical columns (given that $X_{T_1}=X_{T_2}$). Finally, if $\sum_t y_t\neq 1$, we also have $m(y,X;b)=0$.  These last points, combined with \eqref{eq:link_f_mu}-\eqref{eq:for_f2}, imply $f(b)=0$. Thus, $b \in B$ and the result follows.

\subsubsection*{Part 2}

The proof is in two steps. First, we show that for all $b \in R$,
\begin{equation}
\text{sgn}(D_1(X;b)) = - \text{sgn}(D_2(X;b)) \quad \text{a.s.}	
	\label{eq:sign_equality}
\end{equation}
Second, we show that whenever \eqref{eq:sign_equality} holds, we can construct a distribution of $\gamma|X$ such that \eqref{eq:condit_mom} holds. The result then follows.

\medskip
\textbf{First step: \eqref{eq:sign_equality} holds.}

\medskip
First, the result holds for $b=\beta_0$ since then $D_j(X;b)=0$ for $j\in\{1,2\}$. Otherwise, fix $b=c\beta_0\in R$ and let $\tilde \lambda:= (1,c, c\lambda_{2})$ and $\check \lambda:= (\lambda_{2},c, c\lambda_{2})$. Let $p$ (resp. $p'$) denote the minimal number of pairwise coordinate permutations needed to sort the vector $\tilde \lambda$ (resp. $\check \lambda$) and let $\tilde \lambda^s$ (resp. $\check \lambda^s$) be the corresponding vector, sorted in ascending order. If $c \in (1/\lambda_{2}, 1)$, we have $p=1$ and $p'=2$, whereas if $c \in (1, \lambda_{2})$, $p=0$ and $p'=1$. Hence, in all cases, $p'=p+1$.

\medskip
Now, for any $x \in \text{Supp}(X)$, notice that
\begin{align}\label{eq:first_step}
	D_1(x; b, \lambda) & = D_1(x ; \beta_0, \tilde \lambda) = (-1)^p D_1(x; \beta_0, \tilde \lambda^s), \\
	D_2(x; b, \lambda)&  = D_2(x; \beta_0, \check \lambda) =  (-1)^{p'} D_2(x; \beta_0, \check \lambda^s).
\end{align}
Let $p''$ be the minimal number of pairwise coordinates permutations needed to sort the vector $(x_1'\beta_0, x_2'\beta_0, x_3'\beta_0)$ in ascending order, and let $x^s$ denote the corresponding vector, i.e., such that $x_{s1}'\beta_0 \leq x_{s2}'\beta_0\leq x_{s3}'\beta_0$. Then
\begin{align}
	D_1(x; \beta_0, \tilde \lambda^s) & = (-1)^{p''} D_1(x^s; \beta_0, \tilde \lambda^s),  \\
	D_2(x; \beta_0, \check \lambda^s) & = (-1)^{p''} D_2(x^s; \beta_0, \check \lambda^s).  \label{eq:second_step}
\end{align}
Now, by properties of Chebyshev systems, $D_1(x^s; \beta_0, \tilde \lambda^s)$ and $D_2(x^s; \beta_0, \check \lambda^s)$ are both non-negative. Moreover, both are nonzero if and only if $|\{x_1'\beta_0, x_2'\beta_0, x_3'\beta_0\}|=3$. The result follows by  \eqref{eq:first_step}-\eqref{eq:second_step} and $(-1)^p = - (-1)^{p'}$.

\medskip
\textbf{Second step: if \eqref{eq:sign_equality} holds, there exists a distribution of $\gamma|X$ such that \eqref{eq:condit_mom} holds.}

\medskip
Let us define
\begin{equation}
a_i(\gamma,x)=\frac{w_i\exp(\lambda_{i}\gamma)}{\prod_{t=1}^T\left(1 + \sum_{j=1}^{T-1} w_j \exp(\lambda_{j}(x_t'\beta_0 + \gamma)) \right)}.
	\label{eq:def_a_gamma}
\end{equation}
Then, we have
\begin{equation}
	\E[m(Y,X,b)|X=x]= \E\left[a_1(\gamma,x)|X=x\right] D_1(x, b) + \E\left[a_2(\gamma,x)|X=x\right] D_2(x, b). \label{eq:mom_cond_dev} 	
\end{equation}
Hence, if $D_1(x, b) = D_2(x, b) =0$, any distribution of $\gamma|X=x$ satisfies $\E[m(Y,X,b)|X=x]=0$. Now, suppose that $\sgn\left(D_1(x, b)\right) = -\sgn\left(D_2(x, b)\right)\neq 0$. Then $R(x):= - D_1(x, b) / D_2(x, b)>0$. Let us define
$$\gamma_0 := \frac{\ln\left[w_1 R(x)/w_2\right]}{\lambda_{2}-1}.$$
Consider for $\gamma|X=x$ the Dirac distribution at $\gamma_0$. Then, from \eqref{eq:mom_cond_dev}, we obtain that
$\E[m(Y,X,b)|X=x]=0$. The result follows.

\subsection{Theorem~\ref{thm:semiparam}}
	
Let us first summarize the proof. We link the current model with a ``complete'' model where $\gamma$ is also observed. This model is fully parametric and thus can be analyzed easily. Specifically, we show in a first step  that this complete model is differentiable in quadratic mean \citep[see, e.g.][pp.64-65 for a definition]{vandervaart_2000} and has a nonsingular information matrix. In a second step, we establish an abstract expression for the semiparametric efficiency bound. This expression involves in particular the kernel $\mathcal{K}$ of the conditional expectation operator $g \mapsto \E[g(X,Y)|X,\gamma]$. In a third step, we show that
\begin{equation}
	\label{eq:kernel}
	\mathcal{K} = \{(x,y) \mapsto q(x)m(x,y;\beta_0), \E[q^2(X)]<\infty\}.
\end{equation}
The fourth step of the proof concludes.

\paragraph{First step: the complete model is differentiable in quadratic mean and has a nonsingular information matrix.}  
\label{par:first_step}

	Let $p(y|x,g;\beta):=\P(Y=y|X=x,\gamma=g;\beta)$. We check that the conditions of Lemma 7.6 in \cite{vandervaart_2000} hold. Under, Assumptions~\ref{as:basic}-\ref{as:gen_logit}, we have
	$$p(y|x,g;\beta)= \prod_{t:y_t=1}F(x_{t}'\beta + g)\prod_{t:y_t=0}(1 -F(x_{t}'\beta + g)),$$
	where $F$ is $C^{\infty}$ on $\R$ and takes values in $(0,1)$. This implies that $\beta \mapsto \ln p(y|x,g;\beta)$ is differentiable. Let $S_\beta:=\partial \ln p(Y|X,\gamma;\beta)/\partial \beta$ and let $S_{\beta k}$ denote its $k$-th component. We prove that $\E[S_{\beta k}^2]<\infty$. First, remark that
	$$S_{\beta k} = \sum_{t=1}^T \frac{X_{k,t} F'(X_{t}'\beta + \gamma)}{[F(X_{t}'\beta + \gamma)][1-F(X_{t}'\beta + \gamma)]} \left[Y_{t}- F(X_{t}'\beta + \gamma) \right].$$
	Next, we have
	\begin{align}
	| S_{\beta k} | & \leq  \sum_{t=1}^T|X_{k,t}|  \frac{F'(X_t'\beta + \gamma)}{F(X_t'\beta + \gamma)(1-F(X_t'\beta + \gamma))}  \nonumber \\
	& = \sum_{t=1}^T|X_{k,t}| \frac{\sum_{j=1}^{T-1} w_j\lambda_{j} e^{\lambda_{j}(X_t'\beta + \gamma)}}{\sum_{j=1}^{T-1}w_j e^{\lambda_{j}(X_t'\beta + \gamma)}} \nonumber \\
	& \leq \lambda_{T-1} \sum_{t=1}^T|X_{k,t}|, \label{eq:ineg_Z}
	\end{align}
	where we have used the triangle inequality and $|Y_t -F(X_t'\beta + \gamma)|\leq 1$ to obtain the first inequality. Equation \eqref{eq:ineg_Z} and Assumption~\ref{as:basic}.2 imply that $\E[S_{\beta k}^2]<\infty$. By the dominated convergence theorem and again \eqref{eq:ineg_Z}, $\beta\mapsto \E[S_{\beta}S_{\beta}']$ is continuous. Therefore, the conditions in Lemma 7.6 in \cite{vandervaart_2000} hold, and the complete model is differentiable in quadratic mean. Moreover,
$$\E[S_{\beta}S_{\beta}'] = \E[\V(S_{\beta}|X,\gamma)]= \sum_{t=1}^T \E\left[\left(\frac{F'(X_{t}'\beta + \gamma)}{[F(X_{t}'\beta + \gamma)][1-F(X_{t}'\beta + \gamma)]}\right)^2 X_t X_t'\right].$$
Then, if for some $v \in \R^K$, $v' \E[S_{\beta}S_{\beta}']v=0$, we would have $X_t'v=0$ almost surely for all $t\in\{1,\dotsc, T\}$. By Assumption~\ref{as:GMM}.1, this implies $v=0$. Hence, the information matrix $\E[S_{\beta}S_{\beta}']$ is nonsingular.


\paragraph{Second step: $V^\star$ depends on the orthogonal projection of $\E[S_{\beta_0} | X,Y]$ on $\mathcal{K}$.} 
\label{par:second_step_v_star}

Let $\widetilde{\psi}=(\widetilde{\psi}_1,\dotsc,\widetilde{\psi}_K)'$ denote the efficient influence function, as defined p.363 of \cite{vandervaart_2000}. Then $V^\star =\E[\widetilde{\psi} \widetilde{\psi}']$ and $\E[\widetilde{\psi}]=0$. Let $\mathscr{S}:=$span$(S_{\beta_0})$, $\mathscr{G}:= \{q: \E[q^2(X,\gamma)]<\infty, \E[q(X,\gamma)]=0\}$ and for any closed convex set $A$ and any $h=(h_1,\dotsc,h_K)'$, let $\Pi_A$ denote the orthogonal projection on $A$ and $\Pi_A(h)=(\Pi_A(h_1),\dotsc,\Pi_A(h_K))'$. By Equation (25.29), Lemma 25.34 (since the complete model is differentiable in quadratic mean by the first step) and the same reasoning as in Example 25.36 of \cite{vandervaart_2000},  $\widetilde{\psi}$ is the function of $(X,Y)$ of minimal $L^2$-norm satisfying
\begin{equation}
	\label{eq:caract_Psi}
\widetilde{\chi} = \Pi_{\mathscr{S}+\mathscr{G}}(\widetilde{\psi}),
\end{equation}
where $\widetilde{\chi}$ is the efficient influence function of the large model. Because this large model is parametric, we have
\begin{equation}
	\label{eq:expr_chi}
\widetilde{\chi}= \E[S_{\beta_0}S_{\beta_0}']^{-1}S_{\beta_0}.
\end{equation}
Equation \eqref{eq:caract_Psi} implies $\E[(\widetilde{\psi} - \widetilde{\chi})\widetilde{\chi}']=0$. Thus, defining $\ell_{\beta_0}=\E[S_{\beta_0}|Y,X]$, we get
\begin{equation}
	\label{eq:for_Psi}
	\E[\widetilde{\psi} \ell_{\beta_0}']=\E[\widetilde{\psi} S_{\beta_0}']=\text{Id},
\end{equation}
Moreover, because $\E[S_{\beta_0}|X,\gamma]=0$, $\mathscr{S}$ and $\mathscr{G}$ are orthogonal. Thus, \eqref{eq:caract_Psi} is equivalent to $\Pi_\mathscr{S}(\widetilde{\chi}) =\Pi_\mathscr{S}(\widetilde{\psi})$ and $\Pi_\mathscr{G}(\widetilde{\chi}) = \Pi_\mathscr{G}(\widetilde{\psi})$. Moreover, \eqref{eq:expr_chi} implies that $\Pi_\mathscr{G}(\widetilde{\chi})=0$. Hence, $\widetilde{\psi} \in \mathcal{K}^K$.
Now, because $\Pi_{\mathcal{K}}$ is an orthogonal projector, we have
$$\E[\widetilde{\psi} \Pi_{\mathcal{K}}(\ell_{\beta_0})'] = \E[\Pi_{\mathcal{K}}(\widetilde{\psi})\ell_{\beta_0}'] = \E[\widetilde{\psi}\ell'_{\beta_0}] = \text{Id},$$
where the last equality follows by \eqref{eq:for_Psi}. Hence, if $\Pi_{\mathcal{K}}(\ell_{\beta_0})'\lambda=0$ a.s., we would have $\lambda=0$. In other words, $\E[\Pi_{\mathcal{K}}(\ell_{\beta_0})\Pi_{\mathcal{K}}(\ell_{\beta_0})']$ is nonsingular. Now, consider the set
$$ \mathcal{F}  := \left\{\E[\Pi_{\mathcal{K}}(\ell_{\beta_0})\Pi_{\mathcal{K}}(\ell_{\beta_0})']^{-1} \Pi_{\mathcal{K}}(\ell_{\beta_0}) + v:  \; \E[v\Pi_{\mathcal{K}}(\ell_{\beta_0})'] = 0\right\}.$$
$\mathcal{F}$ is thus the set of vector-valued functions $\psi$ satisfying the equation $\E[\psi \Pi_{\mathcal{K}}(\ell_{\beta_0})]=$Id. Hence, $\widetilde{\psi}$ being the element of $\mathcal{F}$ with minimum $L^2$-norm, we obtain
\begin{align*}
\widetilde{\psi} = \E[\Pi_{\mathcal{K}}(\ell_{\beta_0})\Pi_{\mathcal{K}}(\ell_{\beta_0})']^{-1}\Pi_{\mathcal{K}}(\ell_{\beta_0}).
\end{align*}
Finally, because $V^\star =\E[\widetilde{\psi} \widetilde{\psi}']$,
\begin{equation}
	\label{eq:expr_init_Vstar}
V^\star =  \E[\Pi_{\mathcal{K}}(\ell_{\beta_0})\Pi_{\mathcal{K}}(\ell_{\beta_0})']^{-1}.
\end{equation}


\paragraph{Third step: \eqref{eq:kernel} holds.} 
\label{par:third_step_eqref_eq_kernel_holds}

Let $r \in \mathcal{K}$ and let us prove that $r(y,x)=q(x)m(y,x;\beta_0)$ for some $q$. First, by definition of $\mathcal{K}$, we have, for almost all $(g,x) \in \Supp(\gamma,X)$,
	\begin{align}
	0  =& r((0,0,0),x_0) + r((1,0,0), x_0)G(x_1'\beta_0 + g) + r((0,1,0), x_0)G(x_2'\beta_0 + g)  \notag \\
	+ &  r((0,0,1), x_0)G(x_3'\beta_0 + g)  + r((1,1,0), x_0)G(x_1'\beta_0 + g)G(x_2'\beta_0  + g)  \notag \\
	+ & r((1,0,1), x_0)G(x_1'\beta_0 + g)G(x_3'\beta_0+ g) + r((0,1,1), x_0)G(x_2'\beta_0 + g)G(x_3'\beta_0 + g)  \notag \\
	+ & r((1,1,1), x_0)G(x_1'\beta_0 + g)G(x_2'\beta_0 + g)G(x_3'\beta_0 + g). \label{eq:general_eq}
	\end{align}
	Let $a_t \triangleq x_t'\beta_0$ for $t \in \{1,2,3\}$ and, for the sake of conciseness, let us remove the dependence of $r$ on $x$. Then, using Assumption~\ref{as:gen_logit}, we obtain, for almost all $(g,x)$,
	\begin{align*}
	0 =  & A_1 e^{0 \times g} + A_2 e^{ g} + A_3e^{\lambda_{2}g} + A_4e^{2 g}  + A_5e^{2\lambda_{2} g} + A_6e^{(1 + \lambda_{2}) g} + A_7e^{3  g} + A_8 e^{(2+ \lambda_{2}) g} \\
	& + A_9 e^{(1+ 2\lambda_2) g} +A_{10} e^{3\lambda_{2} g},
	\end{align*}
	where
	\begin{align*}
	A_1:= &  r(0,0,0) , \\
	A_2:= &  w_1\left[r(1,0,0)e^{a_1} + r(0,1,0)e^{a_2} + r(0,0,1)e^{a_3}\right] , \\
	A_3 := & w_2\left[r(1,0,0)e^{\lambda_{2}a_1} + r(0,1,0)e^{\lambda_{2}a_2} + r(0,0,1)e^{\lambda_{2}a_3}\right], \\
	A_4 := & w_1^2\left[r(1,1,0)e^{(a_1 + a_2)}+ r(1,0,1)e^{(a_1 + a_3)} +  r(0,1,1)e^{(a_2 + a_3)}\right] , \\
	A_5 := & w_1w_2\left[r(1,1,0)(e^{a_1 + \lambda_{2}a_2}+ e^{a_2 + \lambda_{2}a_1}) + r(1,0,1)(e^{a_1 + \lambda_{2}a_3} + e^{a_3 + \lambda_{2}a_1}) \right. \\
	& \left.+ r(0,1,1)(e^{a_2 + \lambda_{2}a_3} + e^{a_3 + \lambda_{2}a_2})\right], \\
	A_6 := & w_2^2\left[r(1,1,0)e^{\lambda_{2}(a_1 + a_2)}+ r(1,0,1)e^{\lambda_{2}(a_1 + a_3)} +  r(0,1,1)e^{\lambda_{2}(a_2 + a_3)}\right], \\
	A_7 := & w_1^3 r(1,1,1) e^{a_1+a_2+a_3} , \\
	A_8 := & w_1^2 w_2 r(1,1,1) \left[e^{a_1+a_2+\lambda_{2} a_3} + e^{a_1+\lambda_{2} a_2+a_3} + e^{\lambda_{2} a_1+ a_2+a_3}\right], \\
	A_9 := & w_1 w_2^2 r(1,1,1) \left[e^{a_1+\lambda_{2} (a_2+a_3)} + e^{a_2+\lambda_{2} (a_1+a_3)} + e^{a_3 + \lambda_{2} (a_1+ a_2)}\right], \\
	A_{10} := & w_2^3 r(1,1,1)  e^{\lambda_{2}(a_1+a_2+a_3)}.	
	\end{align*}	
	Since $\lambda_{2} = 2$ is excluded by assumption, there are three cases left depending on the number of different exponents in Equation \eqref{eq:general_eq}.
	
\medskip
First, we consider $\lambda_{2} \notin \{3/2, 3\}$. By Lemma~\ref{lem:pol_exp} and because $|\Supp(\gamma|X)|\geq 10$, we obtain $ A_k=0$ for all $k\in\{1,\dotsc,10\}$. $A_1=A_7=0$ imply that $r(0,0,0)=r(1,1,1)=0$. Next, $A_4 = A_6 = 0$ implies that either $r(1,0,1) = r(1,1,0) = r(0,1,1) = 0$ or
\begin{equation}
	\left\{\begin{array}{rcl}
		r(1,1,0) & = & - r(1,0,1)e^{\lambda_{2}(a_3 - a_2)} - r(0,1,1)e^{\lambda_{2}(a_3 - a_1)}, \\[2mm]
		r(1,1,0) & =&  -r(1,0,1)e^{(a_3 - a_2)} - r(0,1,1)e^{(a_3 - a_1)}.
	\end{array}\right.
	\label{eq:general_case1}
\end{equation}
Consider the second case. $A_5=0$ implies, since $(r(1,0,1), r(1,1,0), r(0,1,1)) \neq (0,0,0)$,
$$r(1,1,0) = -r(1,0,1)\frac{e^{a_1 + \lambda_{2}a_3} + e^{a_3 + \lambda_{2}a_1}}{e^{a_1 + \lambda_{2}a_2} + e^{a_2 + \lambda_{2}a_1}} - r(0,1,1)\frac{e^{a_2 + \lambda_{2}a_3} + e^{a_3 + \lambda_{2}a_2}}{e^{a_1 + \lambda_{2}a_2} + e^{a_2 + \lambda_{2}a_1}}.$$
By assumption, for almost every $x=(x_1,x_2,x_3)$, $a_3 \neq a_2$ and $a_3 \neq a_1$. Then, using the latter display with equation \eqref{eq:general_case1} yields, since $\lambda_{2}\neq 1$,
\begin{eqnarray*}
	r(1,0,1) & = & r(0,1,1)\left[e^{\lambda_{2}(a_3 - a_2)} - e^{a_3 - a_2}\right]^{-1}\left[e^{a_3 - a_1} - e^{\lambda_{2}(a_3 - a_1)}\right], \\[2mm]
	r(1,0,1) & = & r(0,1,1)\left[e^{\lambda_{2}(a_3 - a_2)} - \frac{e^{a_1 + \lambda_{2}a_3} + e^{a_3 + \lambda_{2}a_1}}{e^{a_1 + \lambda_{2}a_2}  + e^{a_2 + \lambda_{2}a_1}} \right]^{-1} \\
	& & \times \left[	\frac{e^{a_2 + \lambda_{2}a_3} + e^{a_3 + \lambda_{2}a_2}}{e^{a_1 + \lambda_{2}a_2} + e^{a_2 + \lambda_{2}a_1}} - e^{\lambda_{2}(a_3 - a_1)} \right].
\end{eqnarray*}
Since $(r(1,1,0), r(1,0,1), r(0,1,1)) \neq (0,0,0)$, these equalities and \eqref{eq:general_case1} imply that $r(1,0,1)\neq 0$ and $r(0,1,1) \neq 0$. Then
$$\frac{e^{(1-\lambda_{2})a_2}}{e^{(1-\lambda_{2})a_1}}\frac{e^{a_3 + \lambda_{2}a_2 + (\lambda_{2} - 1)a_1} - e^{\lambda_{2}(a_2 + a_3)}}{e^{\lambda_{2}(a_1+a_2)} - e^{(\lambda_{2}-1)a_2 + \lambda_{2}a_1 + a_3}} = \frac{e^{a_3 + \lambda_{2}a_2 + (\lambda_{2} - 1)a_1} - e^{\lambda_{2}(a_2 + a_3)}}{e^{\lambda_{2}(a_1+a_2)} - e^{(\lambda_{2}-1)a_2 + \lambda_{2}a_1 + a_3}}, $$
which is equivalent to $a_1=a_2$. By assumption, the set of $x$ for which this occurs is of probability zero. In other words, for almost every $x$,
$$r((1,1,0),x) = r((1,0,1),x) = r((0,1,1),x)=0.$$
$A_2 =A_3=0$ implies that either $ r(1,0,0) = r(0,1,0) = r(0,0,1) = 0$ or
$$\left\{\begin{array}{rcl}
	r(0,0,1) & = & - e^{(a_1 - a_3)}r(1,0,0) - e^{(a_2 - a_3)}r(0,1,0), \\
	r(0,0,1) & = & - e^{\lambda_{2}(a_1 - a_3)}r(1,0,0) - e^{\lambda_{2}(a_2 - a_3)}r(0,1,0).	
	\end{array}\right.$$
In the first case, almost surely $r(Y,X)=0 = 0 \times m(Y,X;\beta_0)$. In the second case, $r(Y,X) = q(X) \times m(Y,X;\beta_0) $ for some $q\in L^2_X$. The result follows.

\medskip		
Now, we turn to $\lambda_{2} = 3/2$. Then, for almost all $(g,x) \in \Supp(\gamma,X)$,
\begin{align*}
	0 =  & A_1 e^{0 \times g} + A_2 e^{ g} + A_3e^{\frac{3}{2}g} + A_4e^{2 g}  + (A_5+A_7) e^{3 g} + A_6e^{\frac{5}{2}  g} + A_8 e^{\frac{7}{2} g} + A_9 e^{4 g} + A_{10} e^{\frac{9}{2} g}.
\end{align*}
By Lemma~\ref{lem:pol_exp} and because $|\Supp(\gamma|X)|\geq 9$, we obtain $A_5+A_7=0$ and $A_k=0$ for all $k\not\in\{5,7\}$. $A_1 = A_{10} = 0$ implies that $r(0,0,0) = r(1,1,1) = 0$ which in turn implies that $A_7 =  0$ and thus $A_5=0$. Hence, we have  $A_k=0$ for all $k\in\{1,\dotsc,10\}$ and the same reasoning as when $\lambda_{2} \not\in \{3/2,3\}$ allows us to obtain the result.
		
\medskip
Finally, we consider $\lambda_{2} = 3$. Then, for all $(g,x)$,
$$0 =  A_1 e^{0 \times g} + A_2 e^{ g} + (A_3+A_7)e^{3g} + A_4e^{2 g}  + A_5e^{6 g} + A_6 e^{4 g} + A_7 e^{5 g} + A_8 e^{7 g} + A_9 e^{9 g},$$
By Lemma~\ref{lem:pol_exp} and because $|\Supp(\gamma|X)|\geq 9$, we obtain $A_3+A_7=0$ and $A_k=0$ for all $k\not\in\{3,7\}$. $A_1 = A_{10} = 0$ implies that $r(0,0,0) = r(1,1,1) = 0$ which in turn implies that $A_7 = 0$ and thus $A_3=0$. Hence,  $A_k=0$ for all $k\in\{1,\dotsc,10\}$ and the result follows again as when $\lambda_{2} \not\in \{3/2,3\}$.


\paragraph{Fourth step: conclusion.} 
\label{par:fourth_step_conclusion}

By Steps 2 and 3, there exists $q_0(X)$ such that $\Pi_\mathcal{K}(\ell_{\beta_0}) = q_0(X)m(Y,X;\beta_0)$. Moreover, by definition of the orthogonal projection, $\Pi_\mathcal{K}(\ell_{\beta_0}) - \ell_{\beta_0} \in (\mathcal{K}^\perp)^K$. Hence, again by Step 3, we have, for all $q  \in L^2_X$,
$$\E[q_0(X) q(X) m(Y,X;\beta_0)^2] = \E[\ell_{\beta_0} q(X) m(Y,X;\beta_0)].$$
This implies that
$$q_0(X) \Omega(X)= \E[\ell_{\beta_0} m(Y,X;\beta_0) | X].$$
As a result, because $\ell_{\beta_0}=\E[S_{\beta_0}|Y,X]$,
\begin{align*}
\Pi_\mathcal{K}(\ell_{\beta_0}) = & \Omega^{-1}(X) m(Y,X; \beta_0)\E[\ell_{\beta_0}m(Y,X;\beta_0) | X] \\
= & \Omega^{-1}(X) m(Y,X; \beta_0)\E[S_{\beta_0}m(Y,X;\beta_0) | X].
\end{align*}
Then, using \eqref{eq:expr_init_Vstar}, we obtain
$$V^\star = \E \left[ \Omega^{-1}(X) \E[S_{\beta_0}m(Y,X; \beta_0) | X]\E[S_{\beta_0}m(Y,X; \beta_0) | X]'\right]^{-1}.$$
Now, by the end of the proof of Theorem~\ref{thm:moments}, we have, for all $\beta$,
$$0 = \E_\beta\left[m(Y,X;\beta)|X,\gamma\right].$$
As a result,
	\begin{align*}
	0 & = \nabla_\beta \E_\beta\left[m(Y,X;\beta)|X,\gamma\right] \\
	& = \E_\beta\left[\nabla_\beta m(Y,X;\beta)|X,\gamma\right] +  \E_\beta \left[m(Y,X;\beta) S_{\beta}|X,\gamma\right].
	\end{align*}
	Evaluating this equality at $\beta_0$ and integrating over $\gamma$ yields:
	$$\E[S_{\beta_0}m(Y,X;\beta_0)|X] = - \E[\nabla_{\beta}m(Y,X;\beta_0)|X] = - R(X).$$
	We conclude that
	$$V^\star= \E \left[\Omega^{-1}(X) R(X) R(X)' \right ]^{-1} = V_0,$$
	which is a well-defined matrix by Assumption~\ref{as:GMM}.1.


\end{document}